\documentclass[prb,groupedaddress,preprint,nobibnotes]{revtex4}

\usepackage{graphics}

\begin{document}

\title{Vacancy assisted arsenic diffusion and time dependent clustering effects in 
   silicon}
\date{\today}

\author{Benjamin P. Haley}
\author{Niels Gr{\o}nbech-Jensen}
\affiliation{Department of Applied Science, University of California, 
   Davis CA 95616}

\begin{abstract}
We present results of kinetic lattice Monte Carlo (KLMC) simulations of substitutional 
arsenic diffusion in silicon mediated by lattice vacancies.  Large systems are considered,
with 1000 dopant atoms and long range \textit{ab initio} interactions, to the 18th 
nearest lattice neighbor, and the diffusivity of each defect species over time is calculated.
The concentration of vacancies is greater than equilibrium concentrations in order to 
simulate conditions shortly after ion implantation.
A previously unreported time dependence in the applicability of the pair diffusion model,
even at low temperatures, is demonstrated.  Additionally, long range interactions
are shown to be of critical importance in KLMC simulations;
when shorter interaction ranges are considered only clusters composed entirely of vacancies
form.  An increase in arsenic diffusivity for arsenic concentrations up to 
$10^{19}\ \text{cm}^{-3}$ is observed, along with a decrease in arsenic diffusivity for 
higher arsenic concentrations, due to the formation of arsenic dominated clusters.
Finally, the effect of vacancy concentration on diffusivity and clustering is studied, and
increasing vacancy concentration is found to lead to a greater number of clusters,
more defects per cluster, and a greater vacancy fraction within the clusters.
\end{abstract}


\maketitle

\section{Introduction\label{sec_intro}}
The diffusion of dopants, such as arsenic, in silicon has been studied extensively 
because of its importance in integrated circuit manufacturing.  Native point defects, 
such as lattice vacancies, which may be created during implantation of the dopants, 
interact with other defects to facilitate diffusion of the implanted species.  As 
circuit device sizes continue to shrink, approaching the scale of typical diffusion 
distances, a deeper understanding of this phenomenon is critical.
\par
The effects of temperature on the diffusion process have been studied experimentally and
theoretically.  The experimentally measured diffusivity of arsenic is known to increase 
as temperature increases.\cite{ref_FGP}  Several theoretical models have, as expected, 
demonstrated the same result due to greater thermal fluctuations at higher temperatures.
A previous kinetic lattice Monte Carlo (KLMC) study\cite{ref_BWHJ} showed how defect capture
radii depend on interaction range and temperature.  Bunea and Dunham\cite{ref_BUNEA_PRB} 
used a KLMC model with one arsenic, one vacancy, and interactions ranging from the third to 
sixth nearest lattice neighbor site.  They found an increase in arsenic diffusivity at 
higher temperatures, and quantified deviations from a pair diffusion model in terms of a 
correction factor
introduced into Fick's first law\cite{ref_FICK}. As temperature increased, the value of the
correction factor deviated more from the expected value of 1.  They also demonstrated that
deviations from pair diffusion increased with a longer interaction range.
Pankratov \textit{et al.}\cite{ref_PANK} also used a KLMC model with a single arsenic and 
vacancy and interactions extending from the third to 20th nearest neighbor to show that
arsenic diffusivity increases with temperature.  They demonstrated that a ``ring mechanism",
in which a vacancy moves along a hexagonal ring around an arsenic atom, 
contributes less to arsenic diffusion at higher temperatures and that this effect is more
significant for longer interaction ranges.  In section \ref{sec_temp} we present the
results of similar KLMC calculations, with 1000 arsenic atoms and 100 vacancies and long 
range interactions.  In section \ref{sec_range} we study the effect of interaction range on 
diffusion and clustering.
\par
The effect of arsenic concentration on diffusivity and cluster formation has been studied
because of its relation to the electrical deactivation of arsenic at high doping levels.  
Fair and Weber\cite{ref_FW} experimentally found maximum arsenic diffusivity with arsenic 
concentration of $3 \times 10^{20}\ \text{cm}^{-3}$ at $1000\ ^{\circ}$C.  The diffusivity 
of arsenic decreased for concentrations above and below this level.  
Larsen \textit{et al.}\cite{ref_LARSEN} found that arsenic diffusivity increased for 
arsenic concentrations above $2 \times 10^{20}\ \text{cm}^{-3}$
at $1050\ ^{\circ}$C.  Solmi and Nobili\cite{ref_SN} demonstrated an increase in arsenic 
diffusivity for concentrations up to $3.5 \times 10^{20}\ \text{cm}^{-3}$ and constant 
diffusivity for higher concentrations.
\par
KLMC simulations have also given differing results.  Dunham and Wu\cite{ref_WU} found, 
using estimated third nearest neighbor interactions at $1050\ ^{\circ}$C, an increase 
in arsenic diffusivity at high doping levels, in agreement with 
Larsen \textit{et al.}\cite{ref_LARSEN}.  Bunea and Dunham\cite{ref_BD_MRS} used a 
similar KLMC model, with third nearest neighbor \textit{ab initio} interactions at 
$900\ ^{\circ}$C, and found enhancement over short simulation times, up to $10^4$ time steps,
but the enhancement diminished over longer simulation times.  List and Ryssel\cite{ref_LIST}
also performed KLMC simulations similar to those of Dunham and Wu\cite{ref_WU} but found
no diffusivity enhancement.  All of these KLMC simulations used vacancy concentrations which
depended on the temperature and the equilibrium vacancy concentration, $C_V^0$.  
Experiments\cite{ref_BSYM} performed at $1050\ ^{\circ}$C measured 
$C_V^0 = 3 \times 10^{15}\ \text{cm}^{-3}$.  The value most commonly used in KLMC 
simulations, included those cited above, was $C_V^0 = 5 \times 10^{16}\ \text{cm}^{-3}$,
which is still a dilute concentration but large enough to allow for reasonable system sizes
in KLMC simulations.
\par
Xie and Chen\cite{ref_XC} summarized all these discrepancies, experimental and theoretical, 
and concluded that time dependent clustering effects were responsible.  They performed 
\textit{ab initio}
calculations and showed that interactions to at least the ninth nearest neighbor were 
important and that a short term arsenic diffusivity enhancement would die out due to 
formation of clusters larger than arsenic-vacancy pairs.  
In section \ref{sec_conc}, we present our KLMC results, with long range (18th nearest 
neighbor) \textit{ab initio} interactions, for several arsenic concentrations, and a 
constant arsenic:vacancy ratio.  In this work, we assume a large vacancy concentration, 
which would be expected post-implantation.  In section \ref{sec_ratio} we vary the 
arsenic:vacancy ratio to determine what effect vacancy concentration has on diffusion 
and clustering.

\section{Simulation method}
\subsection{Kinetic Lattice Monte Carlo}
We use a kinetic lattice Monte Carlo (KLMC) method to simulate atomic scale diffusion
processes.  This method visits defects on a silicon lattice, either randomly or in 
some predetermined order, and, if a move is allowed, attempts to move each defect, one 
at a time.  Defects include substitutional dopants and native point defects, which
include vacancies in this work.  Possible movements include vacancies exchanging lattice
sites with lattice atoms or defects.  Although concerted exchange is a theoretical 
possibility\cite{ref_CE}, we do not consider it in this work.  Thus, a move is allowed if
the defect in question is a vacancy, or if the defect in question is a dopant, and 
a randomly chosen neighboring site is occupied by a vacancy.  If a move is attempted,
the change in the system energy determines whether the move is accepted.  If the move
lowers energy, it is accepted.  If the move does not lower energy, it is accepted,
according to the Metropolis detailed balance algorithm, with probability 
$e^{-\Delta E/2k_{B}T}$, where $k_B$ is Boltzmann's constant, $T$ is the system 
temperature, and $\Delta E$ is the change in system energy as a result of the move.
The system energy in each configuration is the sum of pair interactions between the 
defects.  The values for the pair interactions at given 
separations on the lattice, out to 18 neighbors,  were calculated with \textit{ab initio}
methods described in subsection B.
\par
The time scale of a KLMC simulation is set by the hopping frequency, $\nu_H$, of the 
defects, which is determined by 
\begin{equation}\label{eq_hopfreq}\nu_H = \nu_0e^{-E_b/k_BT}\end{equation}
where $\nu_0$ is the attempt frequency, $E_b$ is the height of the energy barrier for
the move, $k_B$ is Boltzmann's constant, and $T$ is the system temperature.  In this work,
for vacancies, we used $\nu_0^V$ calculated with the expression
\begin{equation}\label{eq_attfreq}\nu_0 = \frac{8\text{D}_0}{\text{a}_0^2},\end{equation}
and the values 
$D_0^V = 1.18\times10^{-4}\ \text{cm}^2/\text{s}$,
and $E_b^V = 0.1\ \text{eV}$, reported for low density vacancy diffusion by 
Tang \textit{et al.}\cite{ref_TANG}, and a lattice constant of 5.43 {\AA}.  
Since this value for $E_b$ is less than the 
0.2 eV used by Bunea and Dunham\cite{ref_BUNEA_PRB,ref_BD_MRS}, our vacancy hopping 
frequency is higher.
For the attempt frequency of arsenic, we scaled the attempt frequency of vacancies
\begin{equation}\nu_0^{As} = \sqrt{\frac{m_V}{m_{As}}}\nu_0^V
\label{eq_Asattfreq}\end{equation}
where $m_V$ is the effective mass of a vacancy (or the mass of a silicon atom) and 
$m_{As}$ is the mass of an arsenic atom.  We used 0.58 eV as the energy barrier for 
arsenic--vacancy exchange, as reported by Pankratov \textit{et al.}\cite{ref_PANK}.
\par
After each KLMC step, which visits all defects in the system, the simulation time is 
incremented by the time step of the fastest allowed event, which in this work is the 
exchange of a vacancy with a Si lattice atom, set by $\nu_H^V$.
For events with longer time steps, such as the exchange of a vacancy with an arsenic atom,
an attempt is made at each time step if a randomly generated number between 0 and 1 is 
less than, for example,
\begin{equation}\label{eq_freqratio}
\frac{\nu_H^{As}}{\nu_H^V}.
\end{equation}
\par
KLMC methods follow the position of each defect in the system over time.  This allows the
definition of a diffusion coefficient as
\begin{equation}\label{eq_Dt}
D(t_1,t_2) = \frac{\left\langle\left|r_i(t_2) - r_i(t_1)\right|^2\right\rangle_i}
{6\left|t_2 - t_1\right|},
\end{equation}
which approaches the thermodynamic diffusion coefficient as $\left|t_2 - t_1\right|$ 
approaches infinity.  We use this definition in order to study diffusion during various time
ranges.

\subsection{\textit{Ab initio} Calculations\label{sec_abinit}}
\begin{figure}\scalebox{0.5}[0.5]{\rotatebox{270}{
\includegraphics{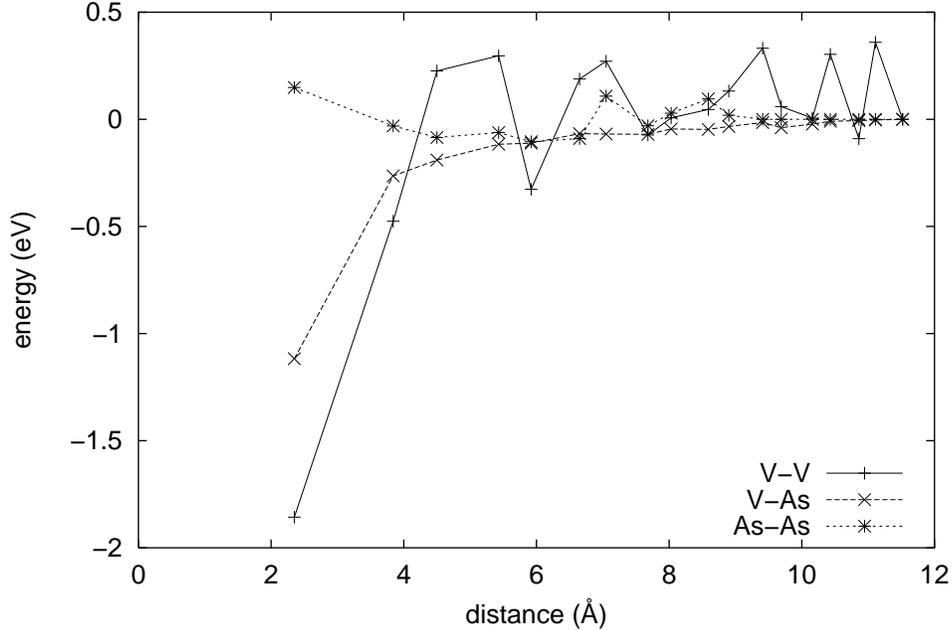}}}
\caption{\label{fig_potentials}
Pair interaction energies for vacancies and arsenic.}
\end{figure}

The pair interactions between vacancies\cite{ref_KMB} and between vacancies and 
arsenic\cite{ref_BWHJ} were calculated using the 
ultrasoft pseudopotential plane wave code VASP\cite{ref_VASP},  
on a 216 atom supercell.  The generalized gradient approximation (GGA) was used, along with 
a $4^3$ Monkhorst-Pack\cite{ref_MPk} \textbf{k}-point sampling and a kinetic energy cutoff of
208 eV.
All vacancies were charge neutral.
The arsenic-arsenic interactions were calculated using PEtot\cite{ref_PEtot_WEB}, a 
norm conserving pseudopotential
plane wave code, on a 64 atom supercell, using the local density approximation (LDA) and a
$2^3$ Monkhorst-Pack \textbf{k}-point sampling and a kinetic energy cutoff of 218 eV.  
In each case, the initial configuration of 
atoms was allowed to relax and the final energy was recorded for all possible separations
of each defect pair.
The results are shown in Figure \ref{fig_potentials}.
In this work, separate defects were not permitted to simultaneously occupy the same lattice 
site.

\section{Results}
For the following sections, we began with a baseline system containing 1000 arsenic atoms 
and 100 vacancies on a silicon lattice measuring 84 unit cells on each side for an 
arsenic concentration of $10^{19}\ \text{cm}^{-3}$.  Boundary conditions were periodic, 
and pair interactions to the 18th nearest neighbor were considered.  The base system 
temperature was 900 K.  
In subsection A, we vary the temperature, while the interaction range varies in
subsection B.  In subsection C  we consider various arsenic 
concentrations, and in subsection D we vary the arsenic:vacancy ratio.
Each simulation presented is the mean of five identical simulations.
We define a cluster as two or more defects, substitutional arsenic or vacancies,
occupying a nearest neighbor position from at least one other member of the cluster.

\subsection{Temperature\label{sec_temp}}
\begin{figure}\scalebox{0.5}[0.5]{\rotatebox{270}{
\includegraphics{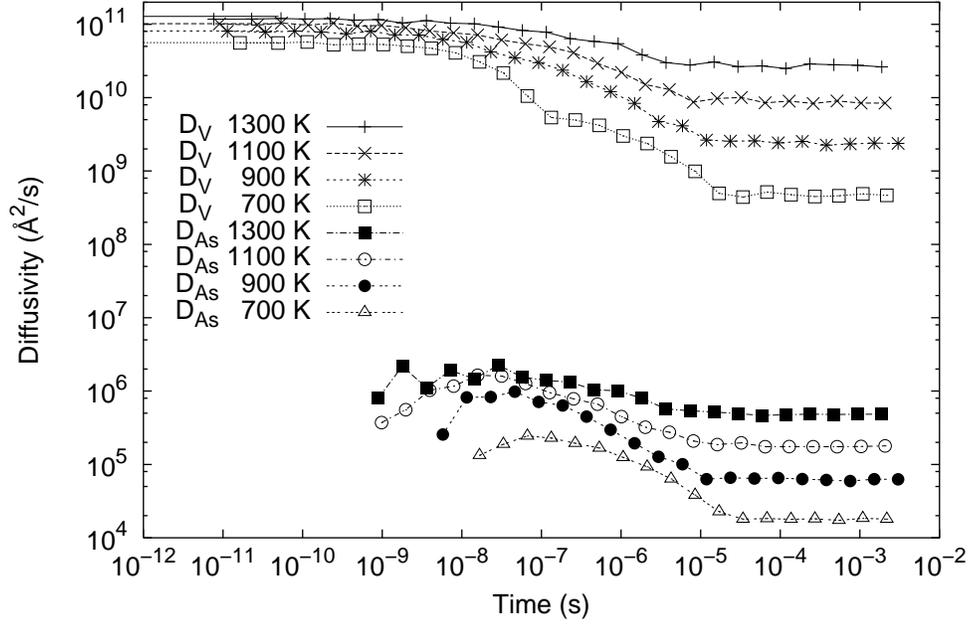}}}
\caption{\label{fig_D_T}
Diffusivity of arsenic and vacancies over time for various temperatures, 
with arsenic concentration $10^{19}\ \text{cm}^{-3}$ and vacancy concentration 
$10^{18}\ \text{cm}^{-3}$ and 18 shell interactions.}
\end{figure}
In this section we vary the system temperature while holding the number of arsenic and 
vacancies, the system size, and the interaction range constant.
Figure \ref{fig_D_T} shows the diffusivity, calculated using equation (\ref{eq_Dt}),
of arsenic and vacancies as a function of time for temperatures ranging from 700 K to 
1300 K.   
The quantity $t_2$ in equation (\ref{eq_Dt}) represents the continuing system time,
and $t_1$ is reset to $t_2$ when $t_2 - t_1$ is twice as large as the previous time interval.
We see that the vacancies diffuse freely at first, while the arsenic atoms are 
stationary.  At all temperatures the free vacancy diffusivities match the theoretical 
values, which are shown on Figure \ref{fig_D_T} for times earlier than the first data point
on the vacancy diffusivity curves.
As arsenic-vacancy pairs (AsV) form, the arsenic begins to diffuse.  The diffusivity of both 
species is greater at higher temperatures, as expected, since the thermal fluctuations
$k_BT$ are larger relative to the migration energy of both AsV and free vacancies,
and by $10^{-5}$ seconds the transient effects disappear.
\begin{figure}\scalebox{0.5}[0.5]{\rotatebox{270}{
\includegraphics{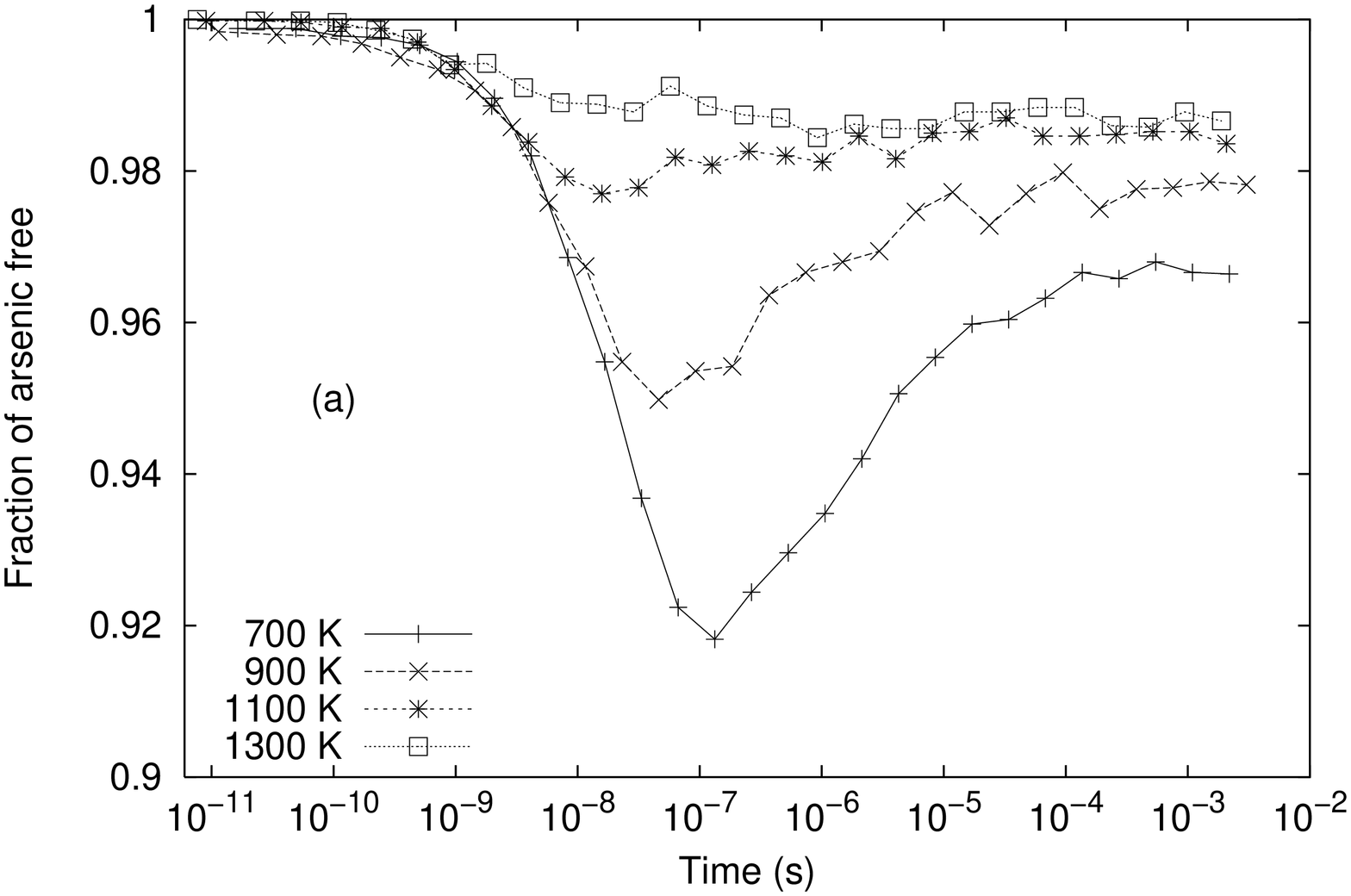}\includegraphics{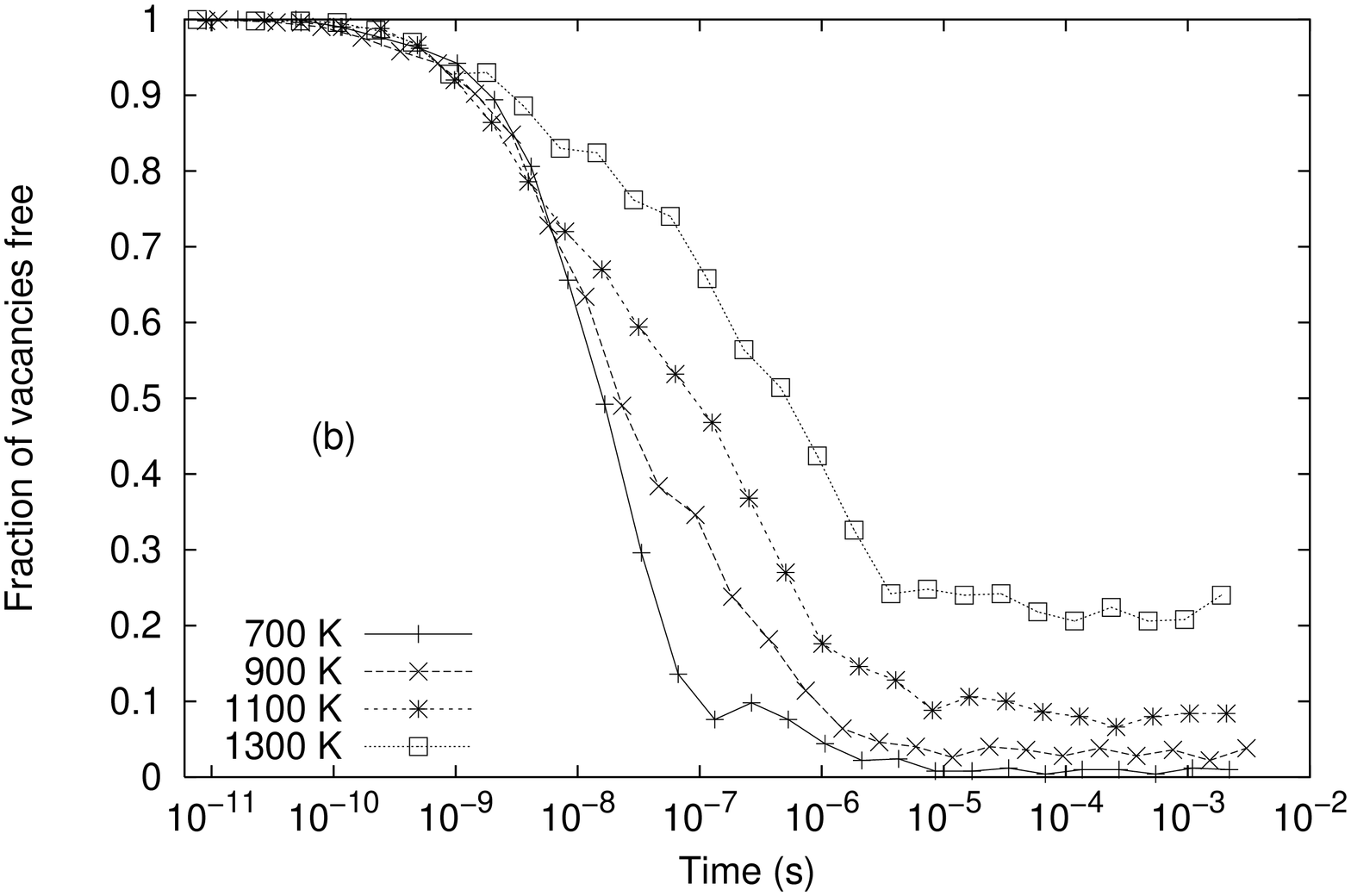}}}
\caption{\label{fig_free_T}
Fraction of arsenic (a) and vacancies (b) free over time for various temperatures,
with arsenic concentration $10^{19}\ \text{cm}^{-3}$ and vacancy concentration 
$10^{18}\ \text{cm}^{-3}$ and 18 shell interactions.}
\end{figure}
\begin{figure}\scalebox{0.5}[0.5]{\rotatebox{270}{
\includegraphics{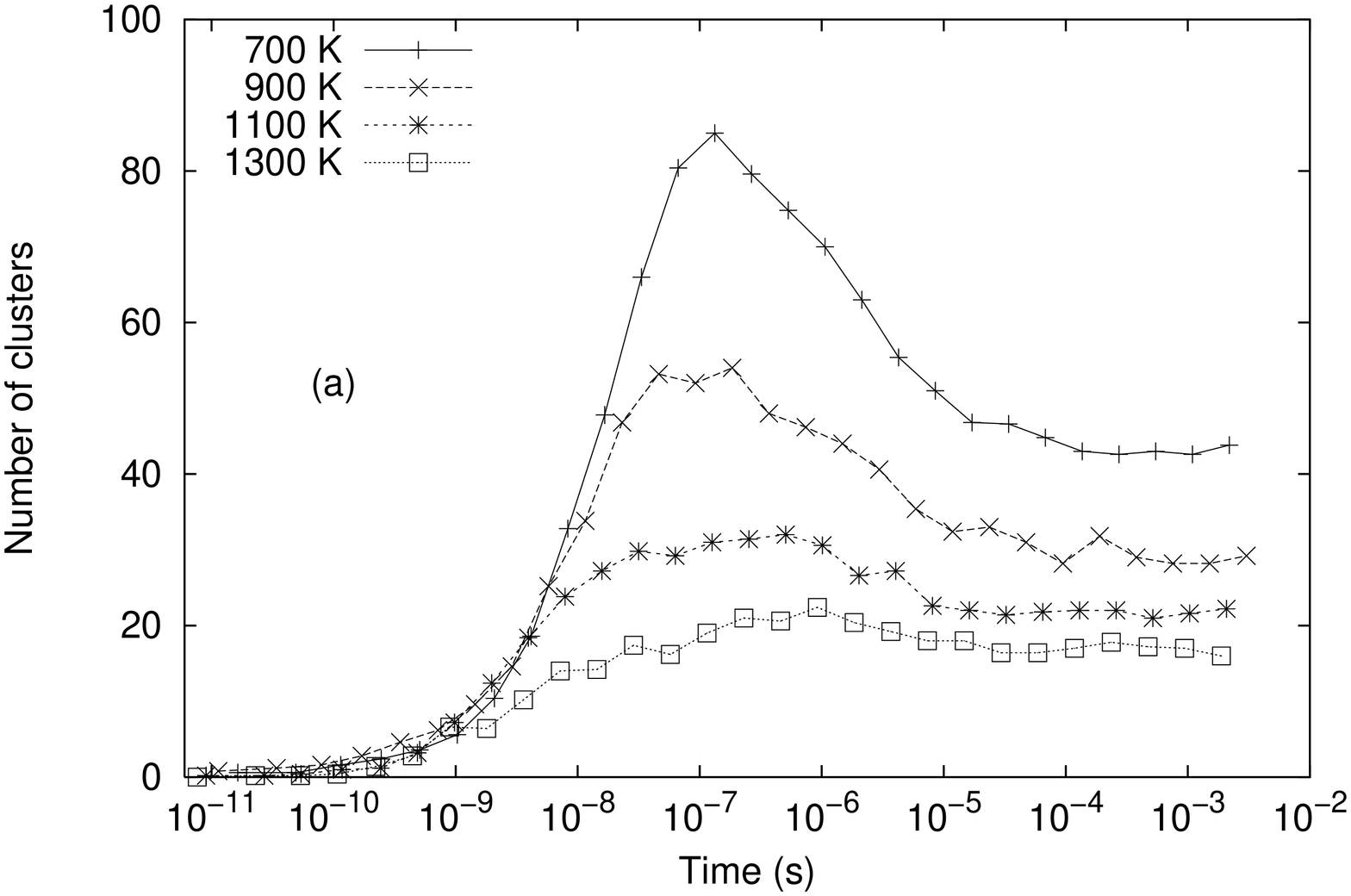}\includegraphics{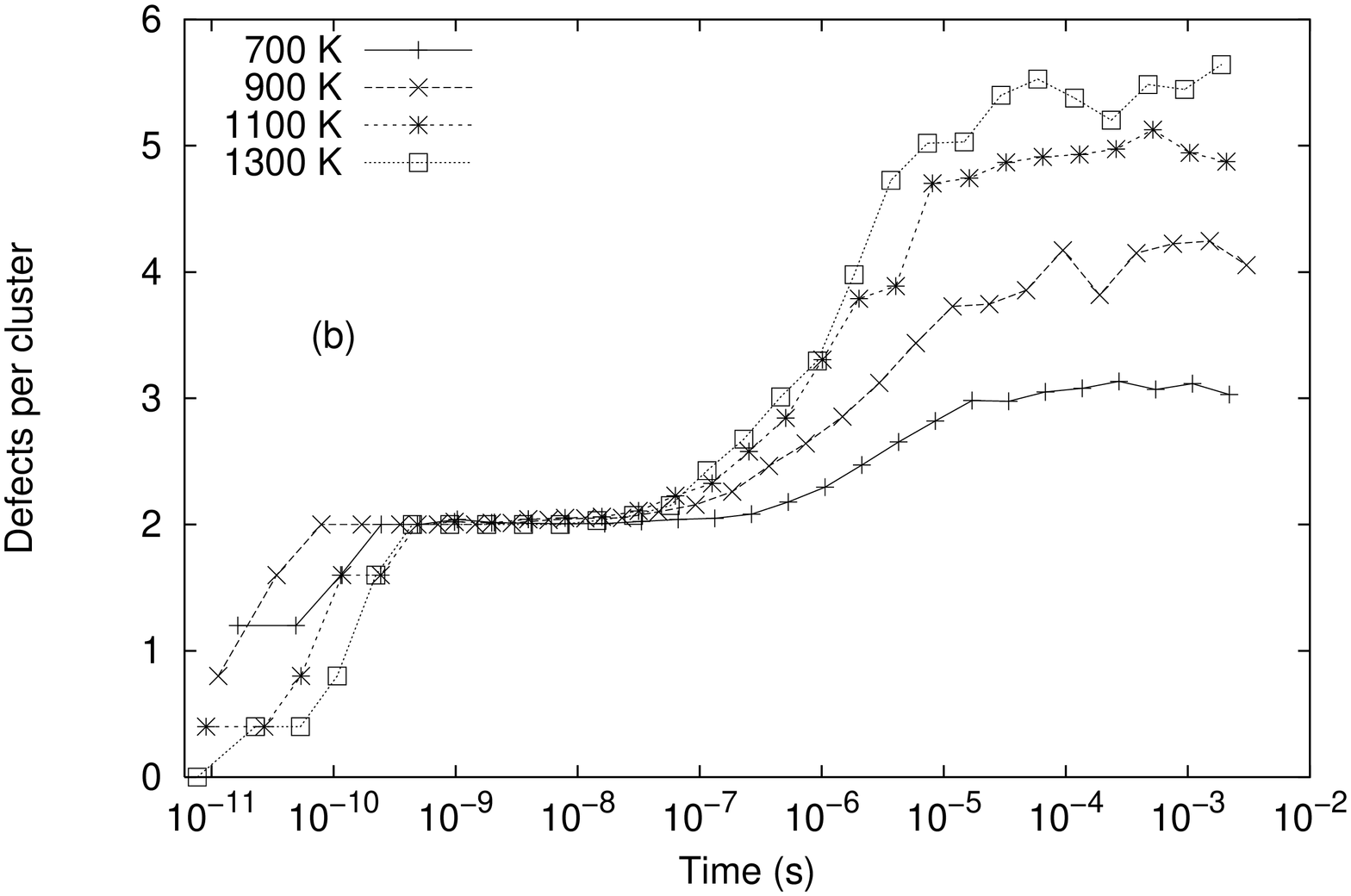}}}
\caption{\label{fig_ndefs_T}
Number of clusters (a) and defects per cluster (b) over time for various temperatures,
with arsenic concentration $10^{19}\ \text{cm}^{-3}$ and vacancy concentration 
$10^{18}\ \text{cm}^{-3}$ and 18 shell interactions.}
\end{figure}
In Figure \ref{fig_free_T}, the fraction of the total number of defects which are free
(not in a cluster) is shown over time for each species.  At higher temperatures, fewer
defects are in clusters.
The same trends can be seen in Figure \ref{fig_ndefs_T}, which shows the number of 
clusters and the number of defects per cluster over time.  More clusters form at lower 
temperatures and these are largely arsenic-vacancy pairs (AsV) for times up to 
approximately $10^{-7}$ seconds.  At that time the number of clusters peaks for lower
temperatures, and the number of defects per cluster rises above two for all temperatures.
At higher temperatures, fewer clusters form, but those that do form are slightly larger on
average.
\begin{figure}\scalebox{0.5}[0.5]{\rotatebox{270}{
\includegraphics{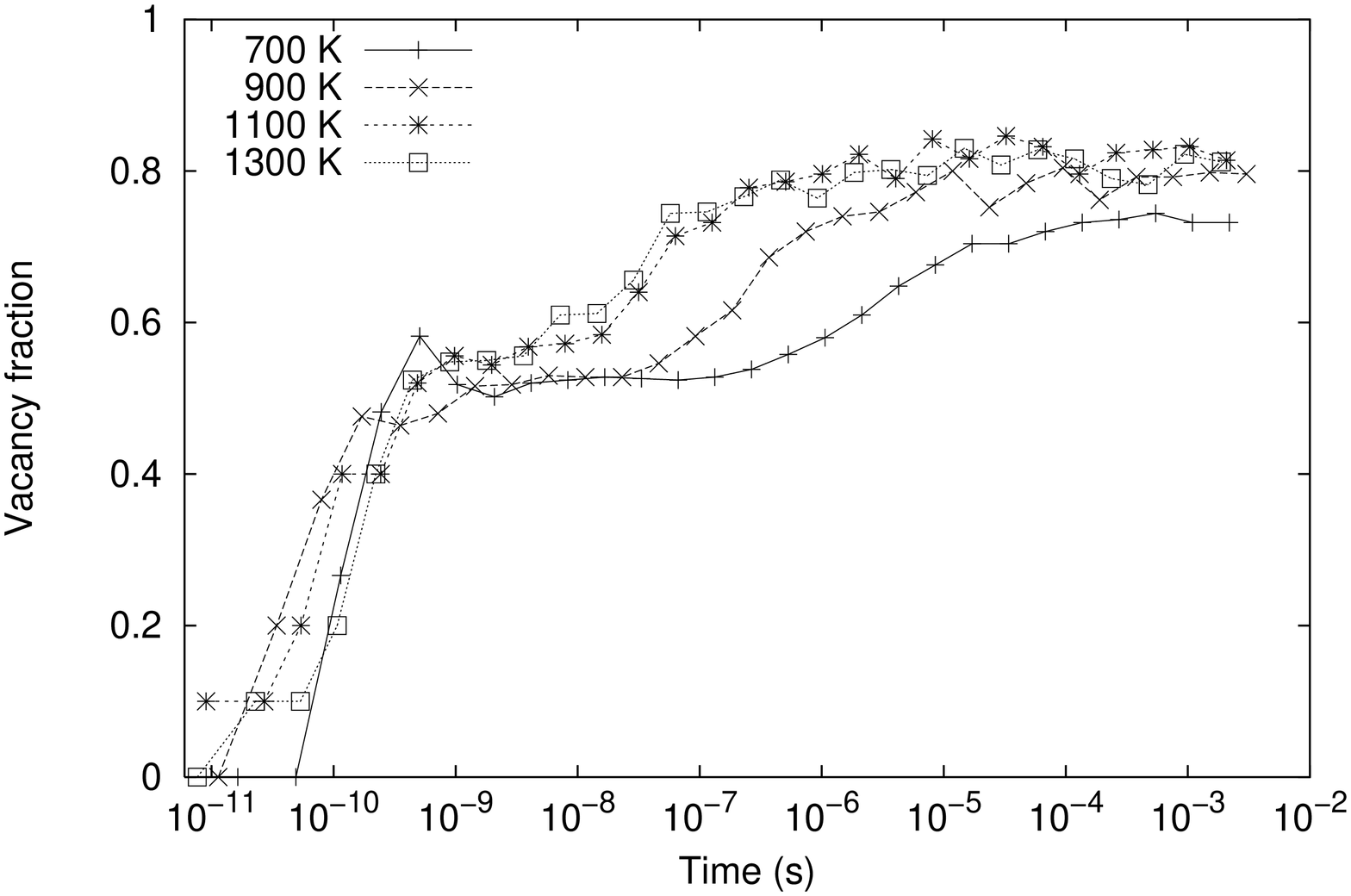}}}
\caption{\label{fig_fractV_T}
Vacancy fraction of defects in clusters over time for various temperatures,
with arsenic concentration $10^{19}\ \text{cm}^{-3}$ and vacancy concentration 
$10^{18}\ \text{cm}^{-3}$ and 18 shell interactions.}
\end{figure}
Figure \ref{fig_fractV_T} shows that the vacancy fraction of defects in clusters increases
over time for all temperatures considered.  For lower temperatures, this implies that 
AsV begin to aggregate, as indicated in Figure \ref{fig_ndefs_T}a, and that some of the
AsV dissociate, as in Figure \ref{fig_free_T}a, with the vacancy attaching itself to 
another AsV.  At higher temperatures the number of clusters is more stable over time,
as in Figure \ref{fig_ndefs_T}a, which implies that mobile AsV attract free vacancies.
In either case, a pair diffusion model is only valid for a short time, roughly $10^{-7}$
seconds.  This result expands the work of Bunea and Dunham\cite{ref_BUNEA_PRB}, who found 
little deviation from the pair diffusion model at lower temperature but did not consider
time dependence.  It should be noted that the results of Bunea and Dunham were obtained
by simulating a single arsenic-vacancy pair with interactions ranging from third to sixth
nearest lattice neighbor.

\subsection{Interaction range\label{sec_range}}
In order to study the effect of interaction range on arsenic diffusion and clustering, we
truncated and shifted the interaction potentials in Figure \ref{fig_potentials} from 18 
shells to 12, 6, and 4 shells, so that the interaction vanishes at the indicated 
separation.  The system temperature, the number of arsenic and vacancies, and the system 
size remained constant.
\begin{figure}\scalebox{0.5}[0.5]{\rotatebox{270}{
\includegraphics{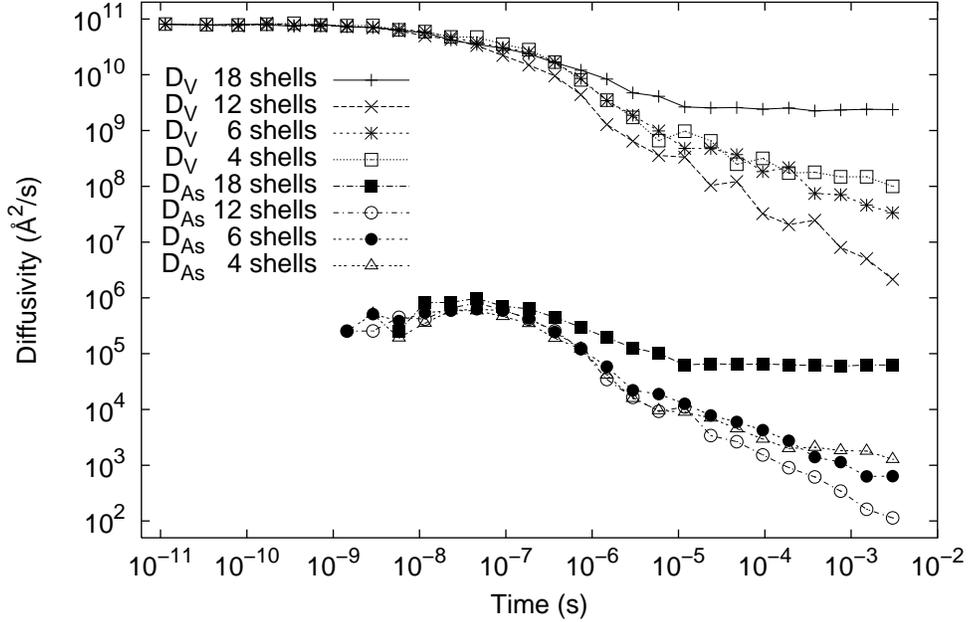}}}
\caption{\label{fig_D_range}
Diffusivity of arsenic and vacancies over time for various interaction ranges,
with arsenic concentration $10^{19}\ \text{cm}^{-3}$ and vacancy concentration 
$10^{18}\ \text{cm}^{-3}$ at 900 K.}
\end{figure}
Figure \ref{fig_D_range} shows the diffusivity of arsenic and vacancies, again calculated
using equation (\ref{eq_Dt}), over time for interaction ranges of 18, 12, 6, and 4 shells.
Those simulations with the longest range interactions show an initial transient behavior
until $10^{-5}$ seconds, after which the diffusivity is constant over time.  In contrast,
for shorter interaction ranges, the diffusivity continues to decrease over time.
\begin{figure}\scalebox{0.5}[0.5]{\rotatebox{270}{
\includegraphics{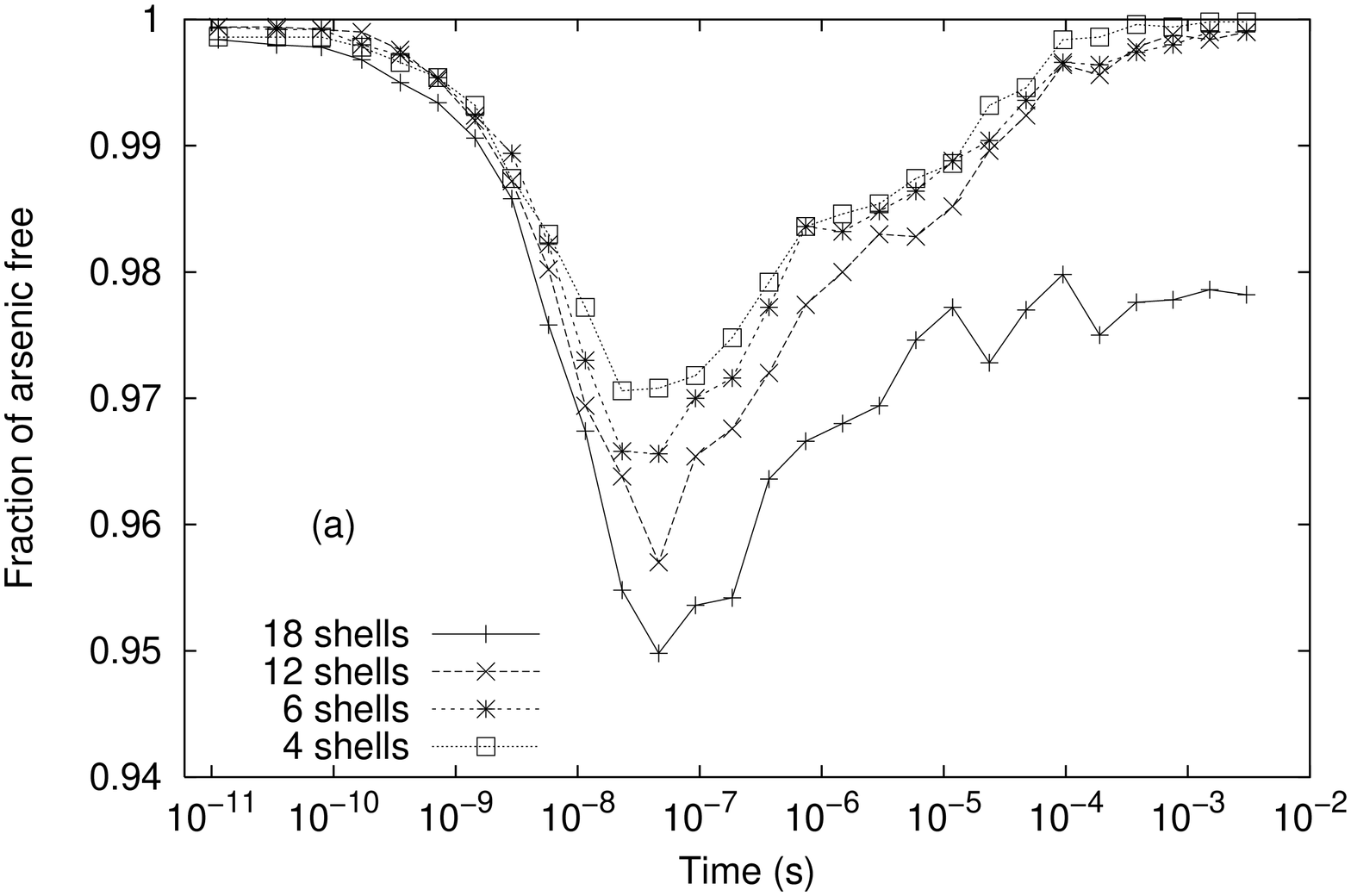}\includegraphics{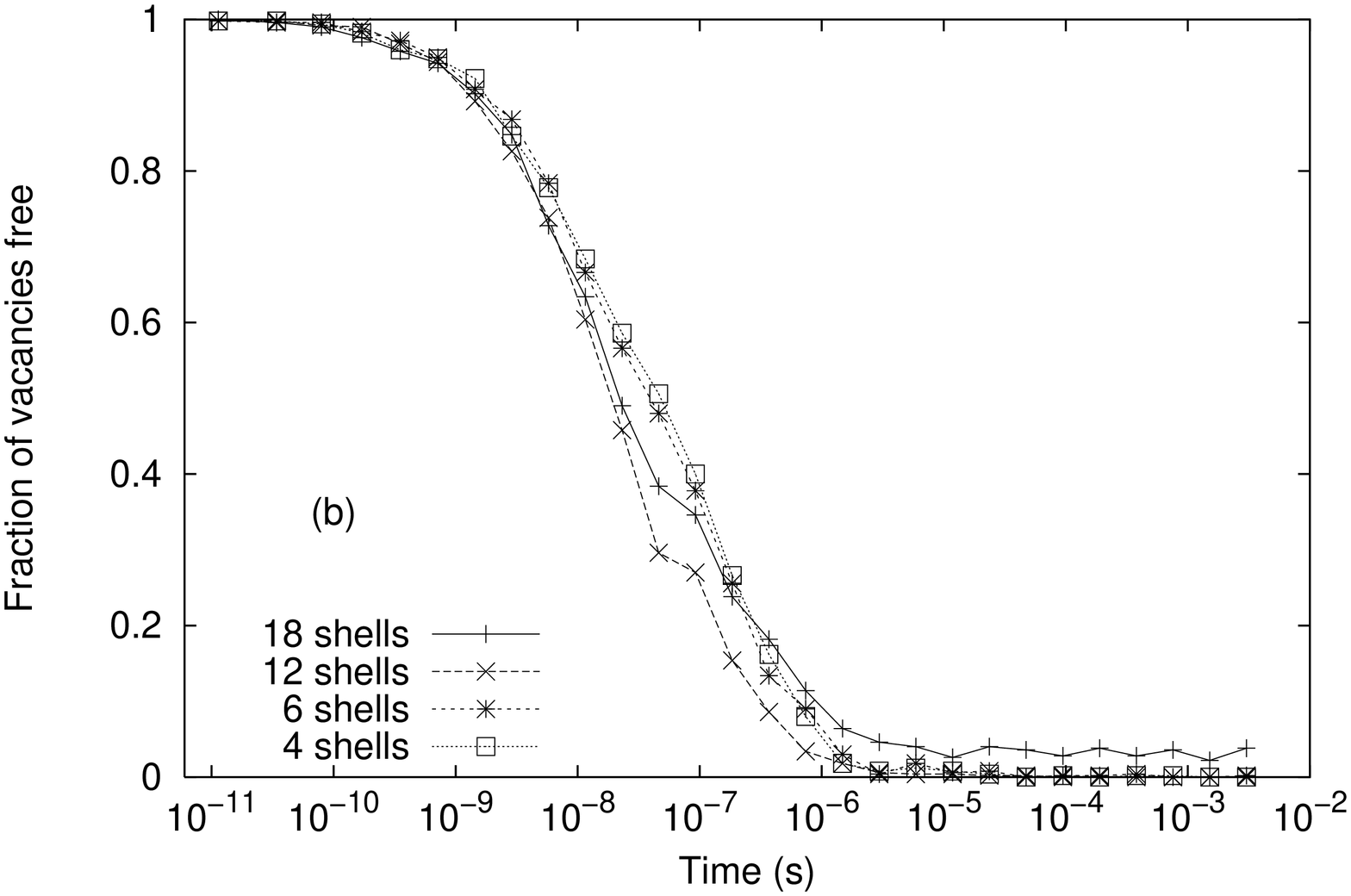}}}
\caption{\label{fig_free_range}
Fraction of arsenic (a) and vacancies (b) free over time for various interaction ranges,
with arsenic concentration $10^{19}\ \text{cm}^{-3}$ and vacancy concentration 
$10^{18}\ \text{cm}^{-3}$ at 900 K.}
\end{figure}
Figure \ref{fig_free_range} shows the fraction of each species free over time.  For all
interaction ranges considered, the fraction of free arsenic decreases as arsenic-vacancy 
pairs (AsV) form, but for 4, 6, and 12 shell interactions, almost all arsenic are free from 
clusters at later times and thus immobile.
For 18 shell interactions, the fraction of arsenic free remains roughly constant over the 
same time range that arsenic diffusivity is constant.  For vacancies, 18 shell interactions
allow some vacancies to remain free at all times, while shorter range interactions lead to 
a state in which all vacancies in the system are bound in clusters.
\begin{figure}\scalebox{0.5}[0.5]{\rotatebox{270}{
\includegraphics{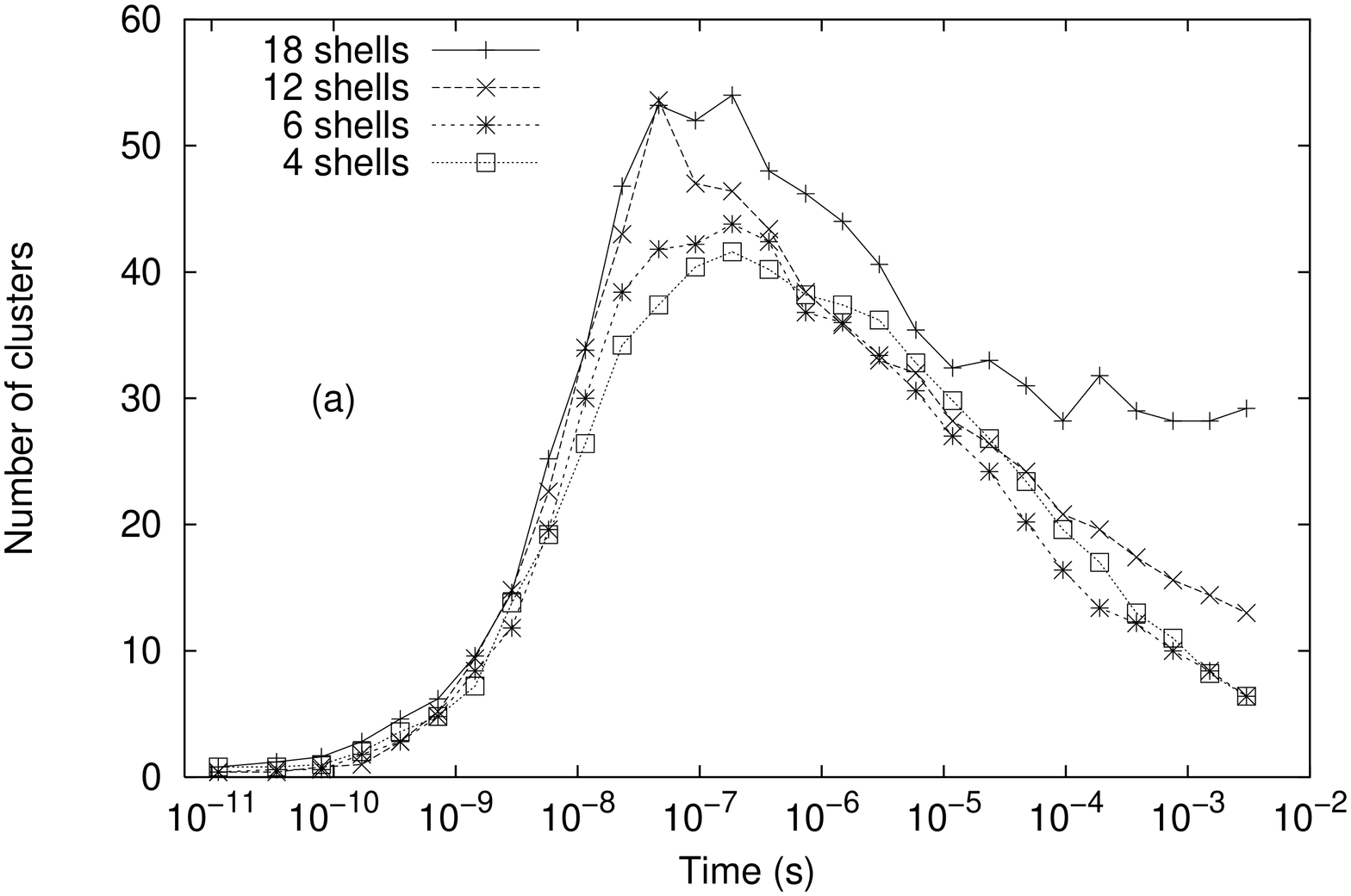}\includegraphics{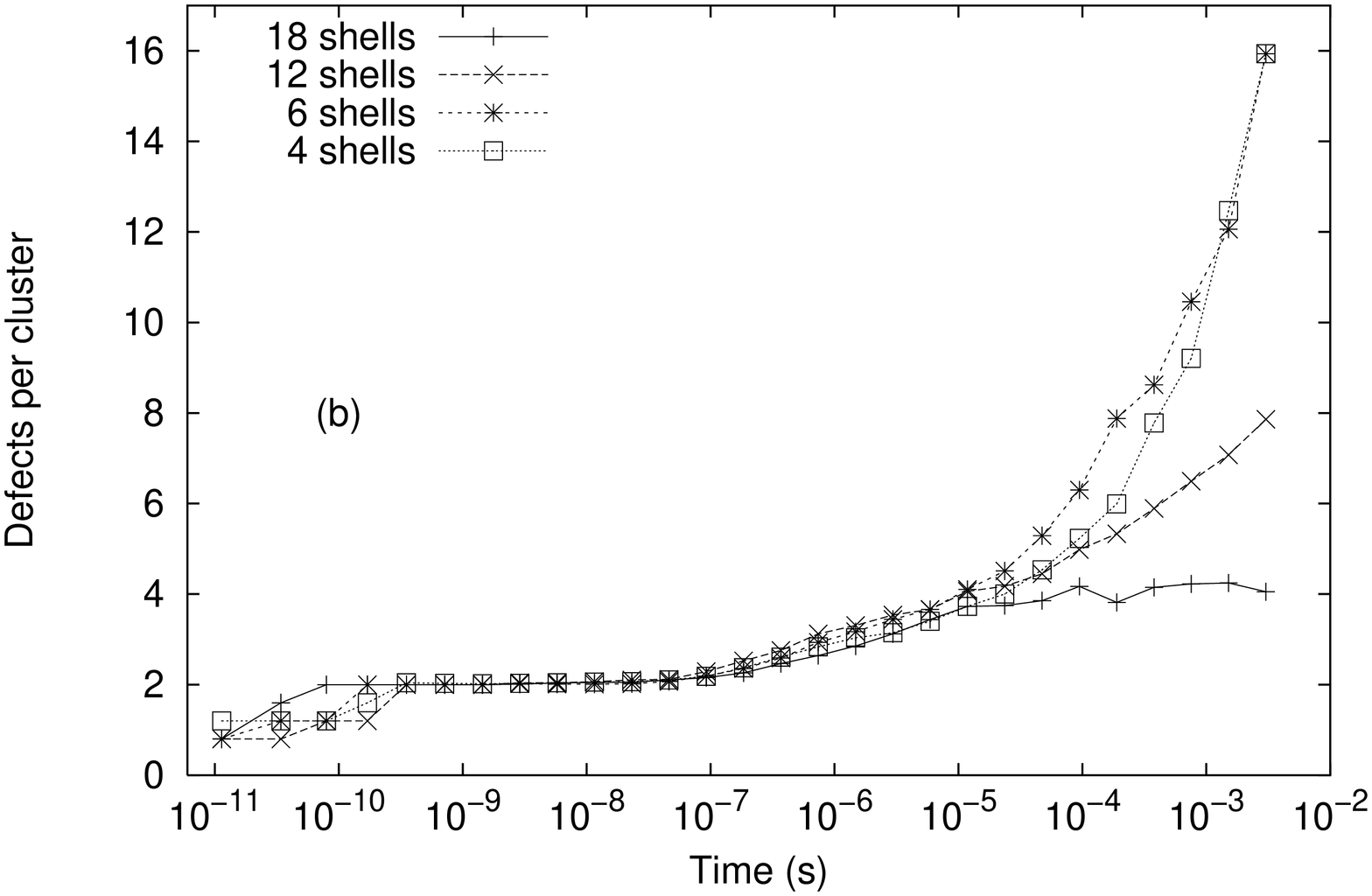}}}
\caption{\label{fig_ndefs_range}
Number of clusters (a) and defects per cluster (b) over time for 
various interaction ranges,
with arsenic concentration $10^{19}\ \text{cm}^{-3}$ and vacancy concentration 
$10^{18}\ \text{cm}^{-3}$ at 900 K.}
\end{figure}
Figure \ref{fig_ndefs_range} presents the number of clusters and defects per cluster over
time for the interaction ranges of interest.  When considering 18 shell interactions,
both quantities reach a somewhat constant level over time, while all shorter range 
interactions cause the number of clusters to continue to decrease and the size of the
clusters to increase.  Combined with the free fractions in Figure \ref{fig_free_range},
this implies that AsV form initially, for times up to $10^{-7}$ seconds, after 
which, as AsV begin to interact, the short range vacancy-vacancy interaction overwhelms
all other interactions; the AsV dissociate, the arsenic atoms are left isolated and
immobile, and the vacancies cluster with each other.
\begin{figure}\scalebox{0.5}[0.5]{\rotatebox{270}{
\includegraphics{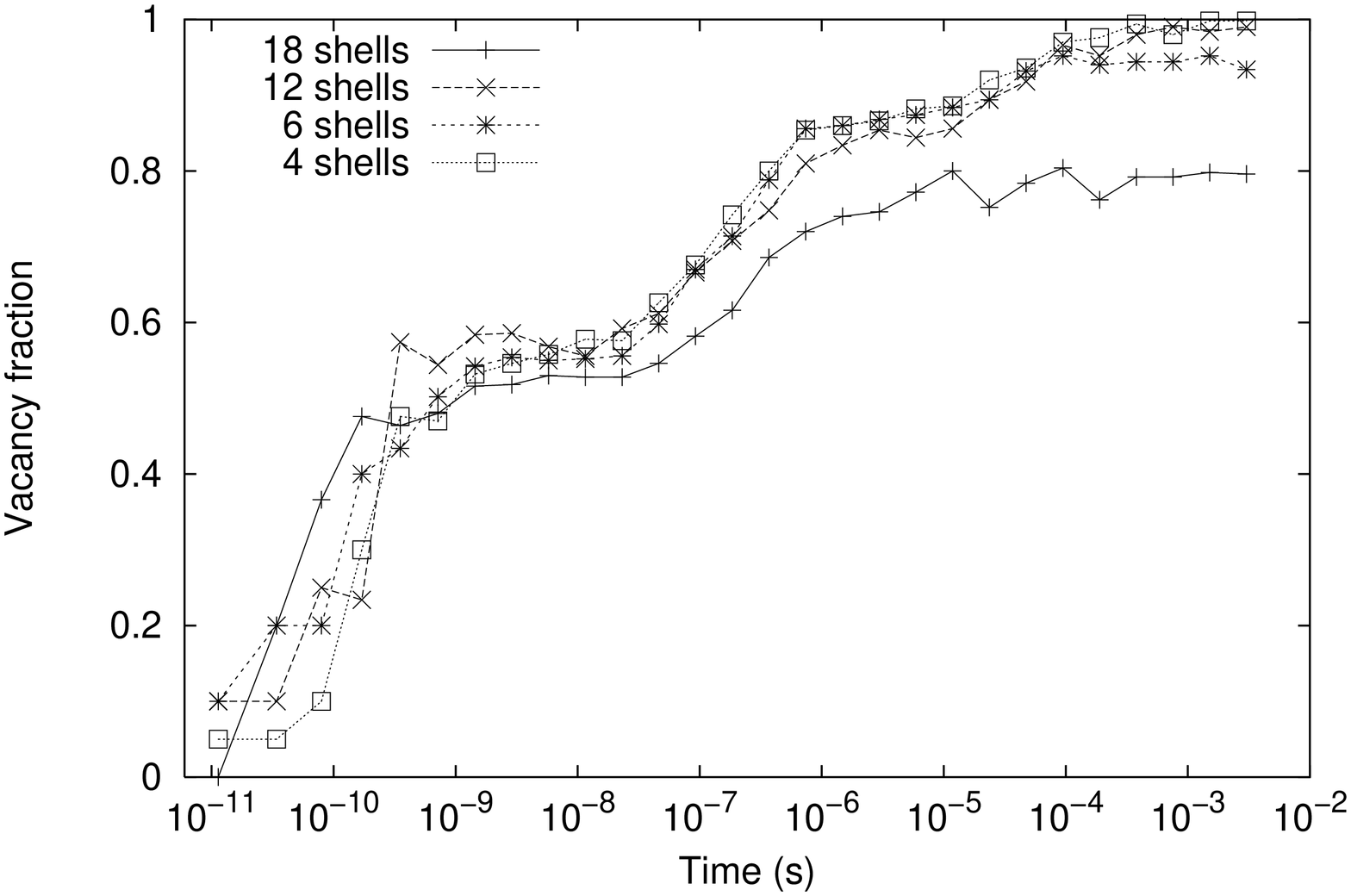}}}
\caption{\label{fig_fractV_range}
Vacancy fraction of defects in clusters over time for various interaction ranges,
with arsenic concentration $10^{19}\ \text{cm}^{-3}$ and vacancy concentration 
$10^{18}\ \text{cm}^{-3}$ at 900 K.}
\end{figure}
This can be seen more clearly in Figure \ref{fig_fractV_range}; for shorter interaction
ranges, almost all defects in clusters are vacancies at long times.  The strong 
interaction between vacancies at short range dominates the clustering behavior of the
system if no longer-range interactions are included.

\subsection{Concentration\label{sec_conc}}
\begin{figure}\scalebox{0.5}[0.5]{\rotatebox{270}{
\includegraphics{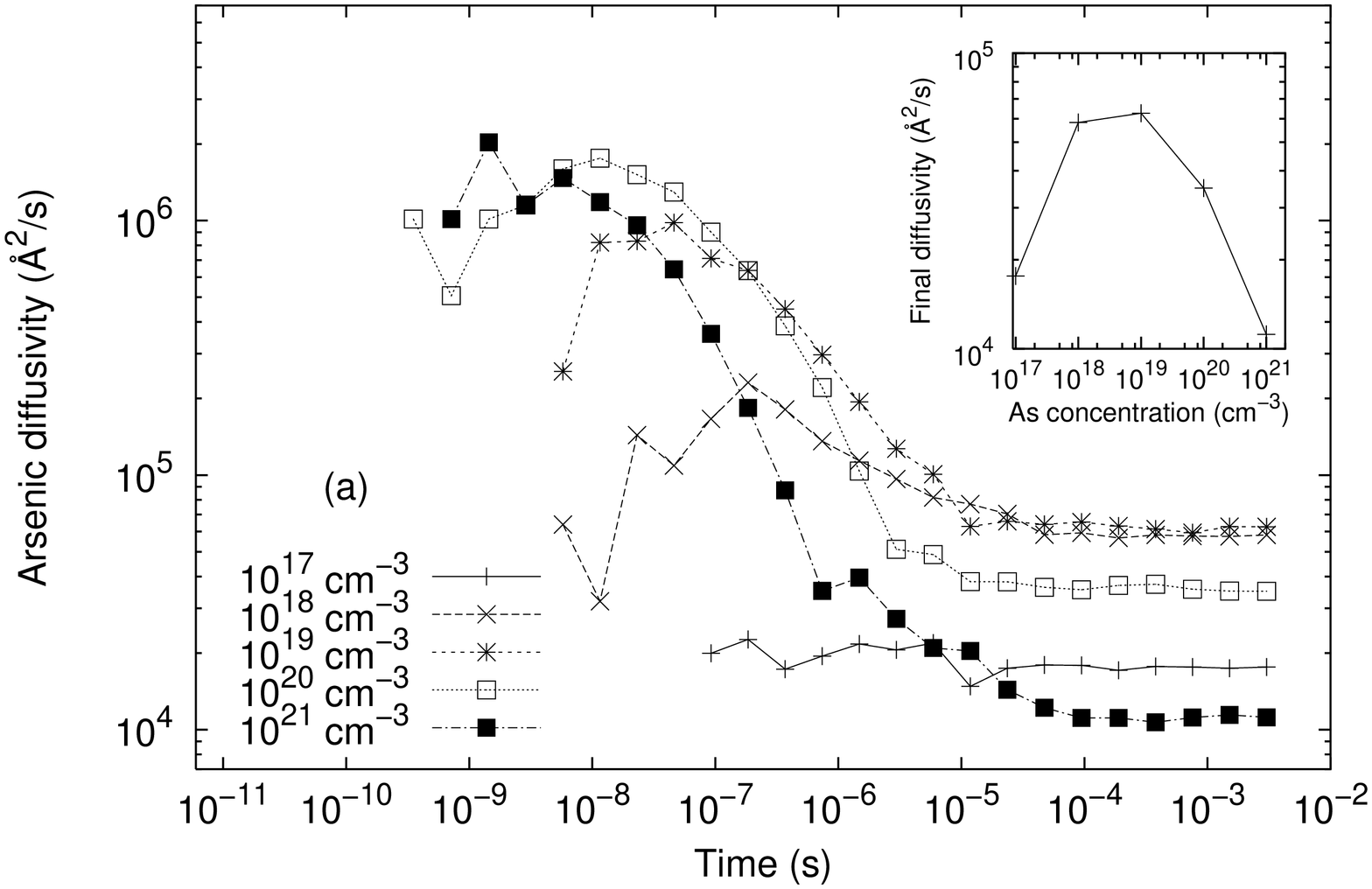}\includegraphics{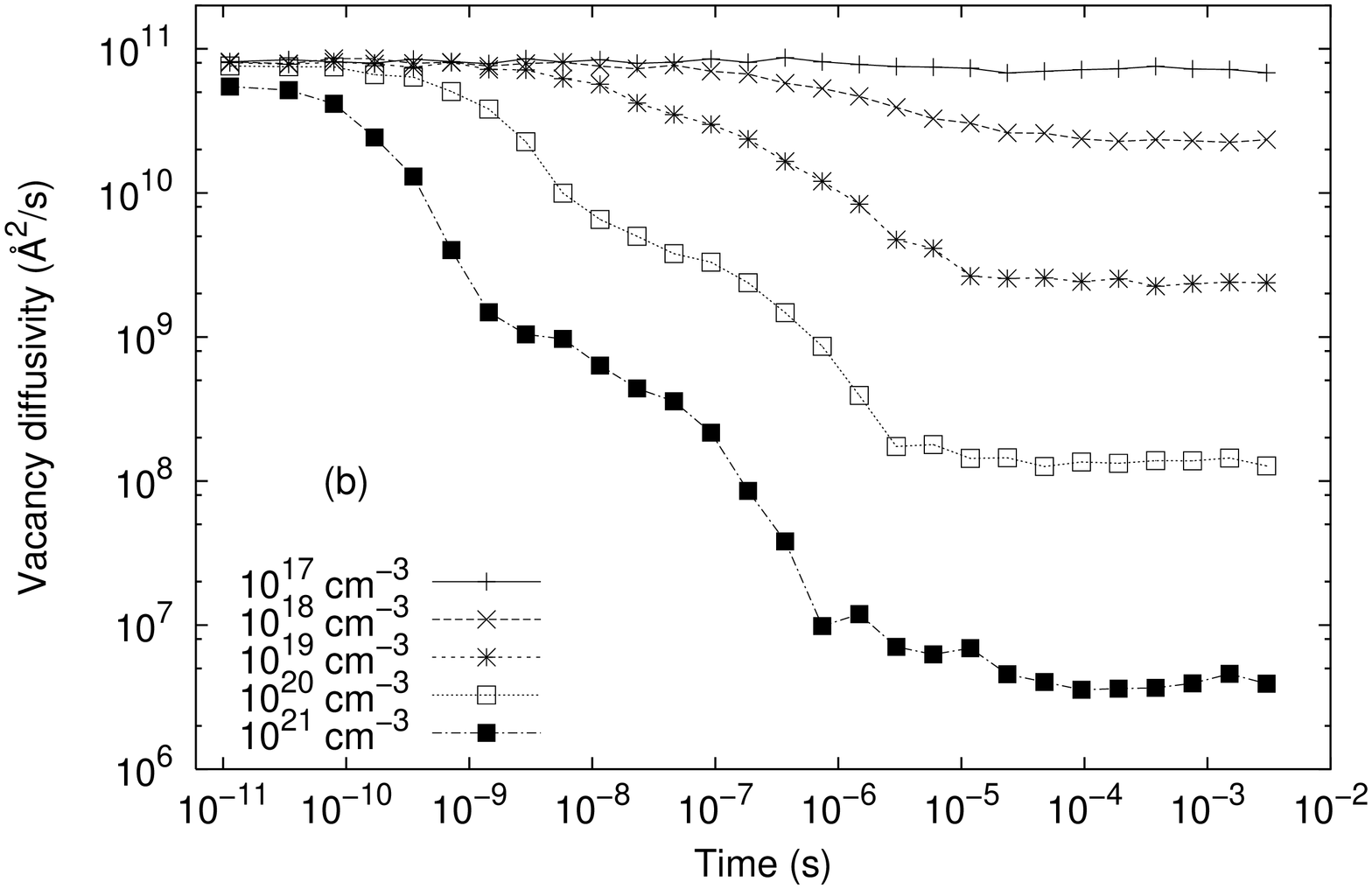}}}
\caption{\label{fig_D_C}
Diffusivity of arsenic (a) and vacancies (b) over time for various arsenic concentrations,
with arsenic:vacancy ratio of 10:1 and 18 shell interactions at 900 K.}
\end{figure}
The concentration of dopants in the system can be varied by two methods.  Either the 
dimensions of a simulation cell with periodic boundary conditions or the number of dopants
in the cell can change.
The diffusivities for both arsenic and vacancies, calculated using equation (\ref{eq_Dt}), 
are virtually identical over time for each method at high concentrations.  At low 
concentrations, some time dependent differences emerge in diffuse systems.  
We chose to vary arsenic concentration by changing the size of the simulation cell, with 
periodic boundary conditions, maintaining a constant number of arsenic and vacancies in the 
system.  This method is much more computationally efficient, as the simulation time
increases with the number of defects in the system.  The system temperature and the 
interaction range are constant, as well.
\par
Figure \ref{fig_D_C} shows the diffusivity, calculated using equation (\ref{eq_Dt}), of 
arsenic and vacancies over time for arsenic concentrations ranging from 
$10^{17}\ \text{cm}^{-3}$ to $10^{21}\ \text{cm}^{-3}$.
The arsenic diffusion begins as arsenic-vacancy pairs (AsV) form, which happens earlier 
as the arsenic concentration increases.  At the highest arsenic concentration considered, 
$10^{21}\ \text{cm}^{-3}$,
there is almost no free diffusion of vacancies, due to the long range
interactions.  Figure \ref{fig_D_C}a shows that, while arsenic diffusivity is initially 
higher at greater arsenic concentrations, by $10^{-5}$ seconds, the maximum arsenic 
diffusivity occurs for concentrations in the range of 
$10^{18}\ \text{cm}^{-3}$ to $10^{19}\ \text{cm}^{-3}$, while
concentrations greater and less than this range yield a lower diffusivity.  This is the
same trend observed experimentally by Fair and Weber\cite{ref_FW}, although the peak 
diffusivity measured in that work was at $3 \times 10^{20}$ As cm$^{-3}$, at a temperature
of $1000\ ^{\circ}$C.  The results in Figure \ref{fig_D_C}a show a different trend than the 
experimental work of Larsen \textit{et al.}\cite{ref_LARSEN} and the KLMC results of 
Dunham and Wu\cite{ref_WU}.  Both of those studies, conducted at $1050\ ^{\circ}$C, 
found an increase in arsenic diffusivity for dopant concentrations above 
$2 \times\ 10^{20}\ \text{cm}^{-3}$.  The simulations of Dunham and Wu
considered third nearest neighbor interactions which decreased linearly with distance.  Our
results also disagree with those of Bunea and Dunham\cite{ref_BD_MRS},  who found greater 
enhancement in arsenic diffusivity at higher arsenic concentrations for short simulation 
times at $900\ ^{\circ}$C; the effect they reported decreased over long simulation times.  
It should be noted that the simulations of Bunea and Dunham considered 
interactions ranging from the third to sixth nearest lattice neighbor, in contrast
to our long range interactions.
\par
\begin{figure}\scalebox{0.5}[0.5]{\rotatebox{270}{
\includegraphics{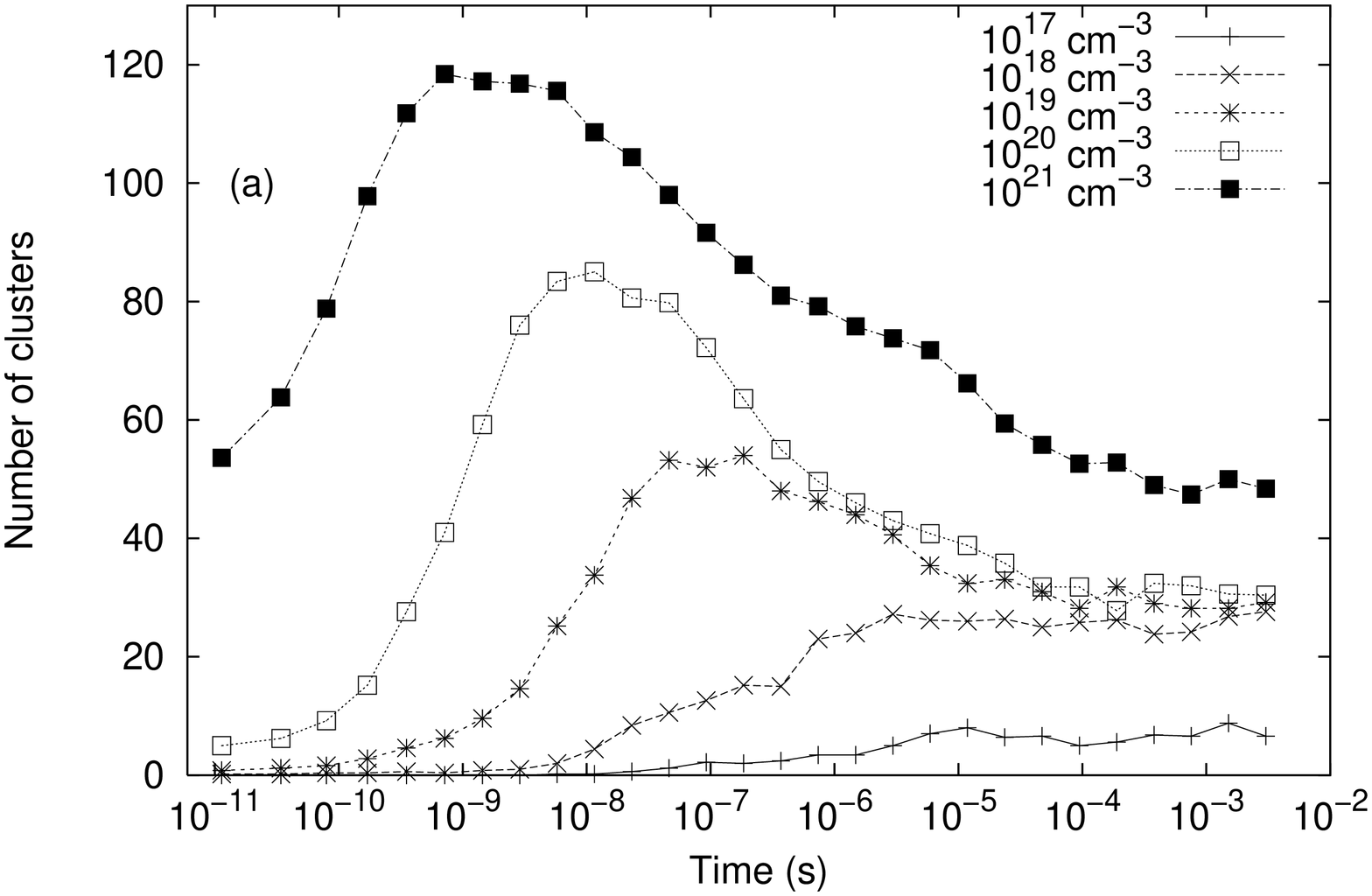}\includegraphics{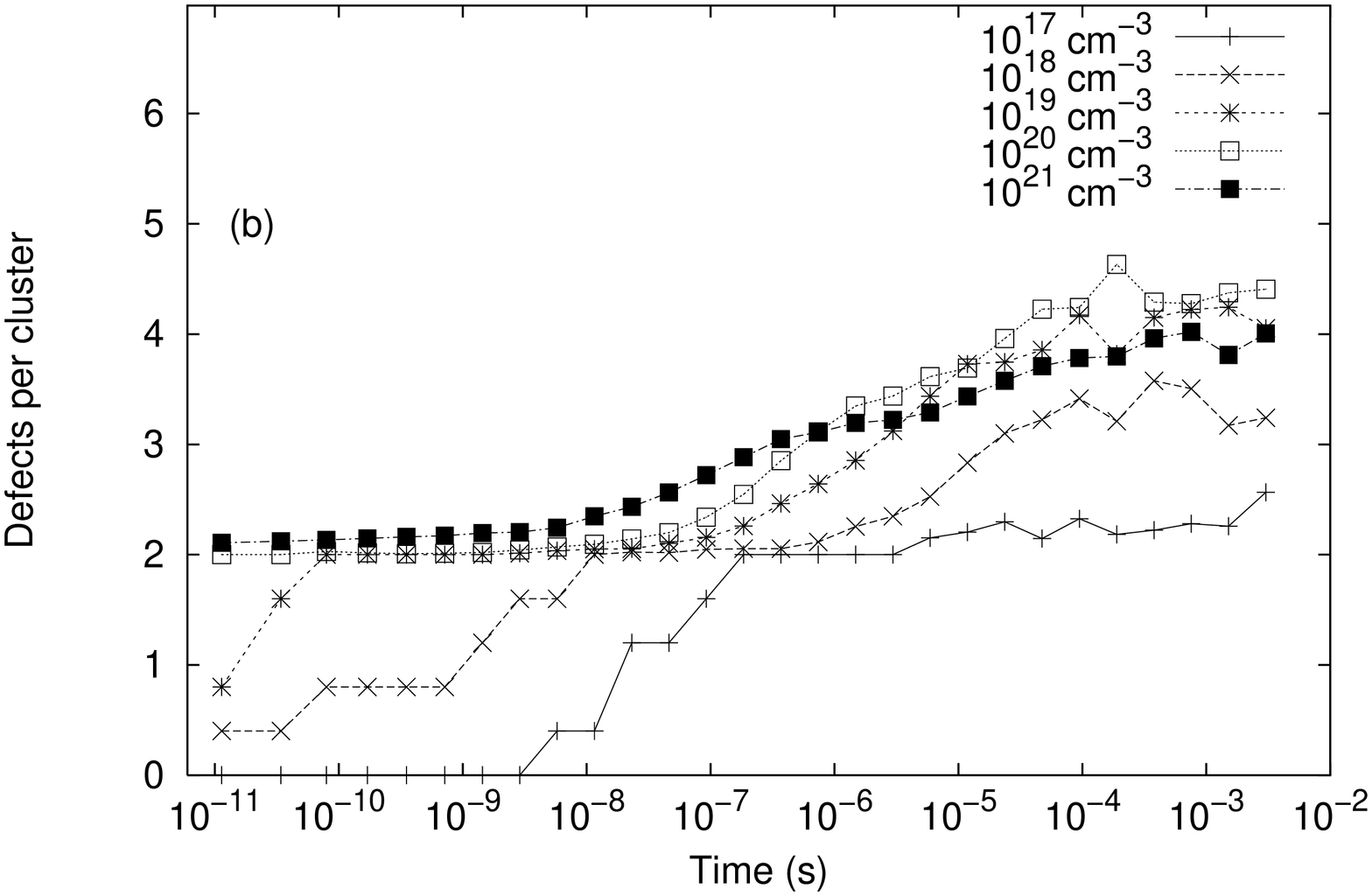}}}
\caption{\label{fig_nclusters_C}
Number of clusters (a) and defects per cluster (b) over time for various
arsenic concentrations,
with arsenic:vacancy ratio of 10:1 and 18 shell interactions at 900 K.}
\end{figure}
\begin{figure}\scalebox{0.5}[0.5]{\rotatebox{270}{
\includegraphics{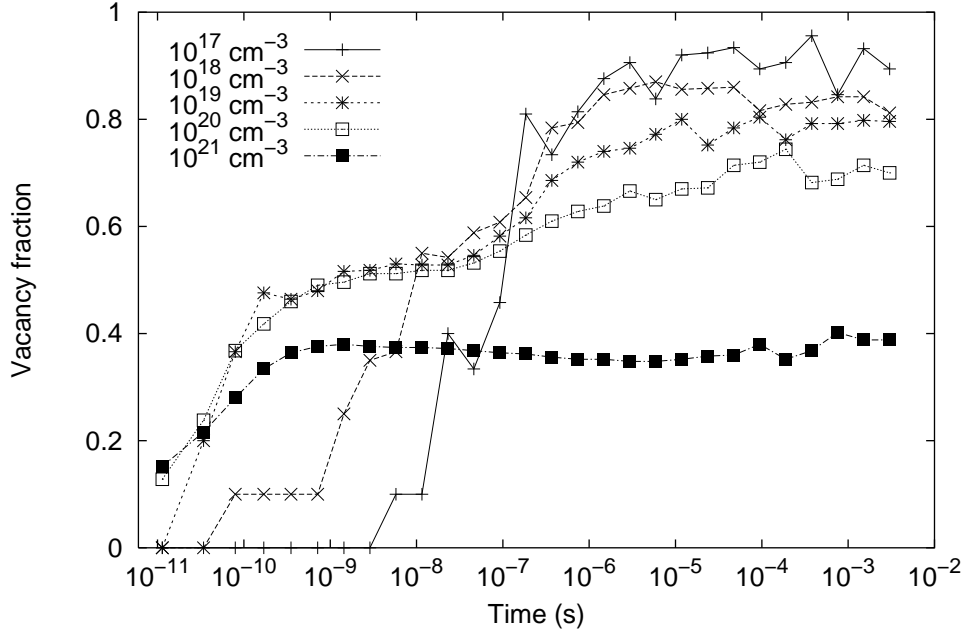}}}
\caption{\label{fig_fractV_C}
Vacancy fraction of defects in clusters over time for various arsenic concentrations,
with arsenic:vacancy ratio of 10:1 and 18 shell interactions at 900 K.}
\end{figure}

Figure \ref{fig_nclusters_C} shows that the number of clusters over time is greater as
arsenic concentration increases, and that the typical cluster size is roughly the same for
different doping levels.   This alone does not explain why arsenic diffusivity is not 
enhanced at greater concentrations, but Figure \ref{fig_fractV_C} shows that as the 
arsenic concentration increases, the fraction of vacancies in clusters decreases.  Thus, 
at higher doping levels the arsenic fraction of defects in clusters increases.  This 
supports the conclusions of Xie and Chen\cite{ref_XC} and Bunea and Dunham\cite{ref_BD_MRS} 
who suggested the development of large, arsenic dominated clusters at elevated doping 
levels.

\subsection{Defect ratio\label{sec_ratio}}
Immediately after ion implantation, the concentration of vacancies should be much higher than 
the equilibrium concentration.  We have assumed a baseline vacancy concentration of 
$10^{18}\ \text{cm}^{-3}$ in this work, for an arsenic:vacancy ratio of 10:1.  In order to 
explore the ramifications of this choice, we performed KLMC simulations at several 
different arsenic:vacancy ratios, ranging from 2:1 to 200:1, with a constant arsenic 
concentration of $10^{19}\ \text{cm}^{-3}$.  The arsenic:vacancy ratio of 200:1 
corresponds to the equilibrium
vacancy concentration used in previously cited KLMC works.  The system temperature, the 
interaction range, and the size of the simulation cell were held constant.
\begin{figure}\scalebox{0.5}[0.5]{\rotatebox{270}{
\includegraphics{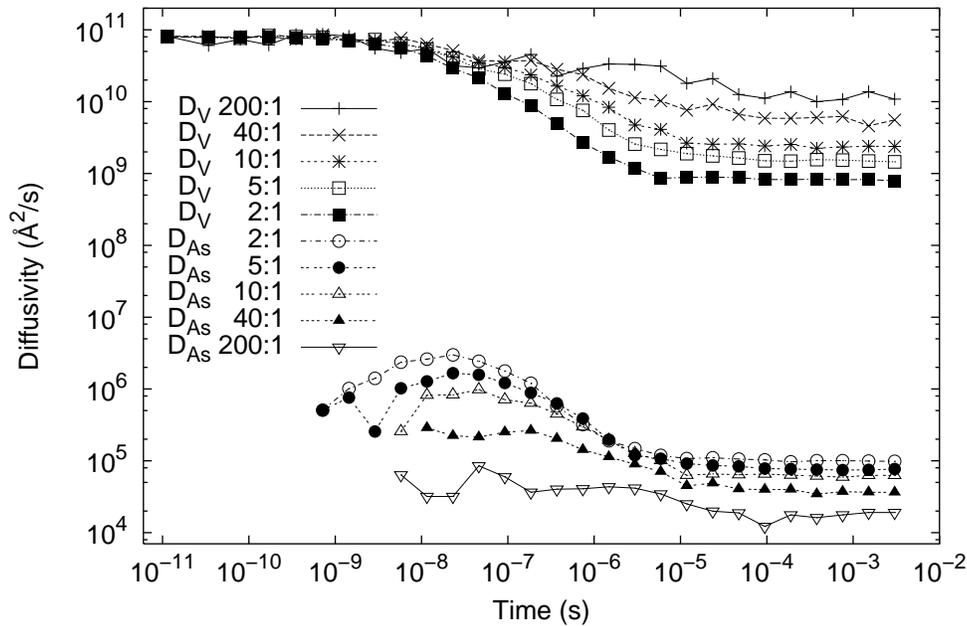}}}
\caption{\label{fig_D_ratio}
Diffusivity of arsenic and vacancies over time for various arsenic:vacancy ratios,
with arsenic:vacancy ratio of 10:1 and 18 shell interactions at 900 K.}
\end{figure}
Figure \ref{fig_D_ratio} shows the diffusivities of arsenic and vacancies, calculated using
equation (\ref{eq_Dt}), for various arsenic:vacancy ratios.  
The diffusivity of arsenic is 
greater with more vacancies in the system.  Conversely, vacancy diffusivity is lower with 
more vacancies in the system.
\begin{figure}\scalebox{0.5}[0.5]{\rotatebox{270}{
\includegraphics{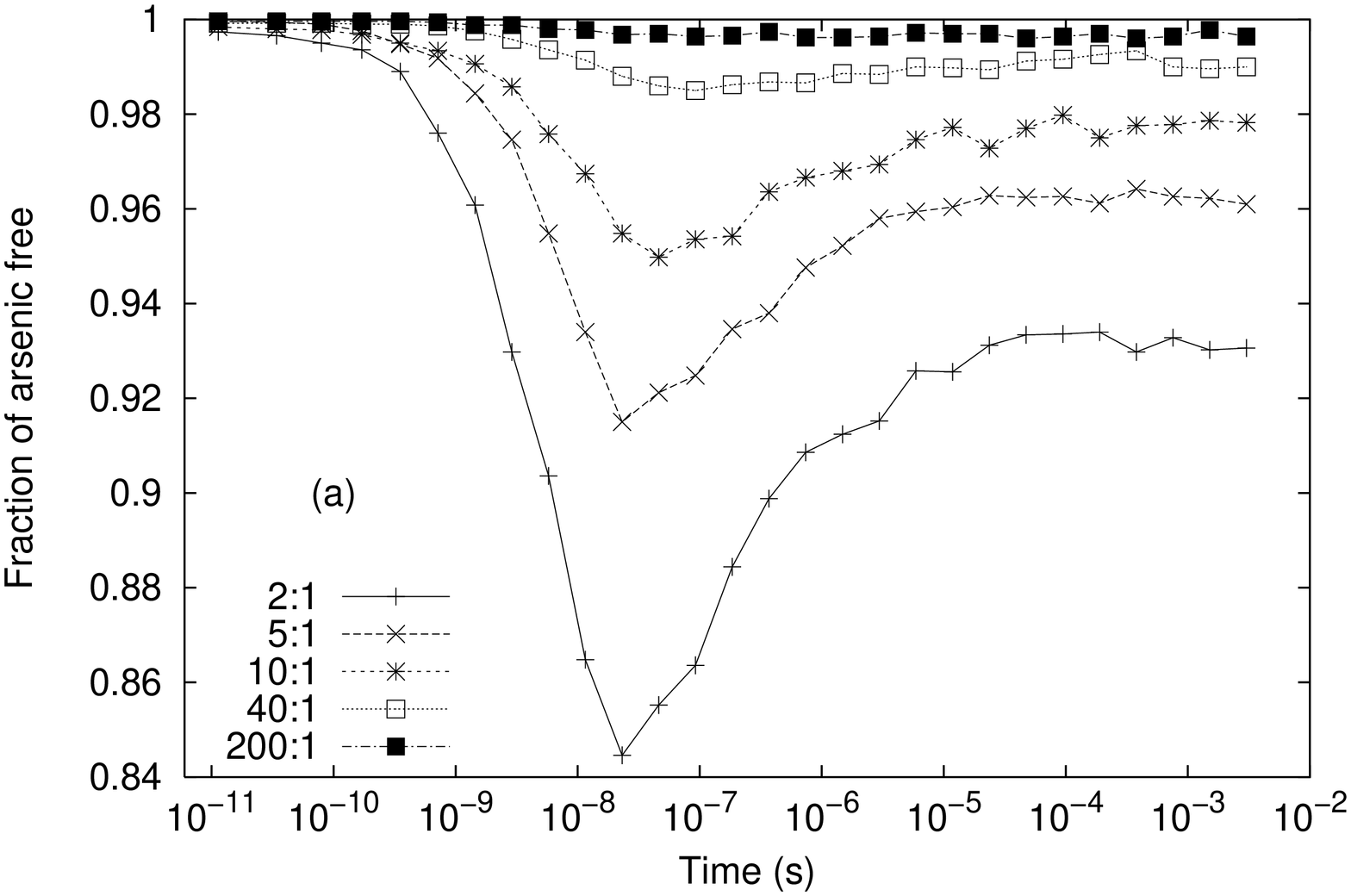}\includegraphics{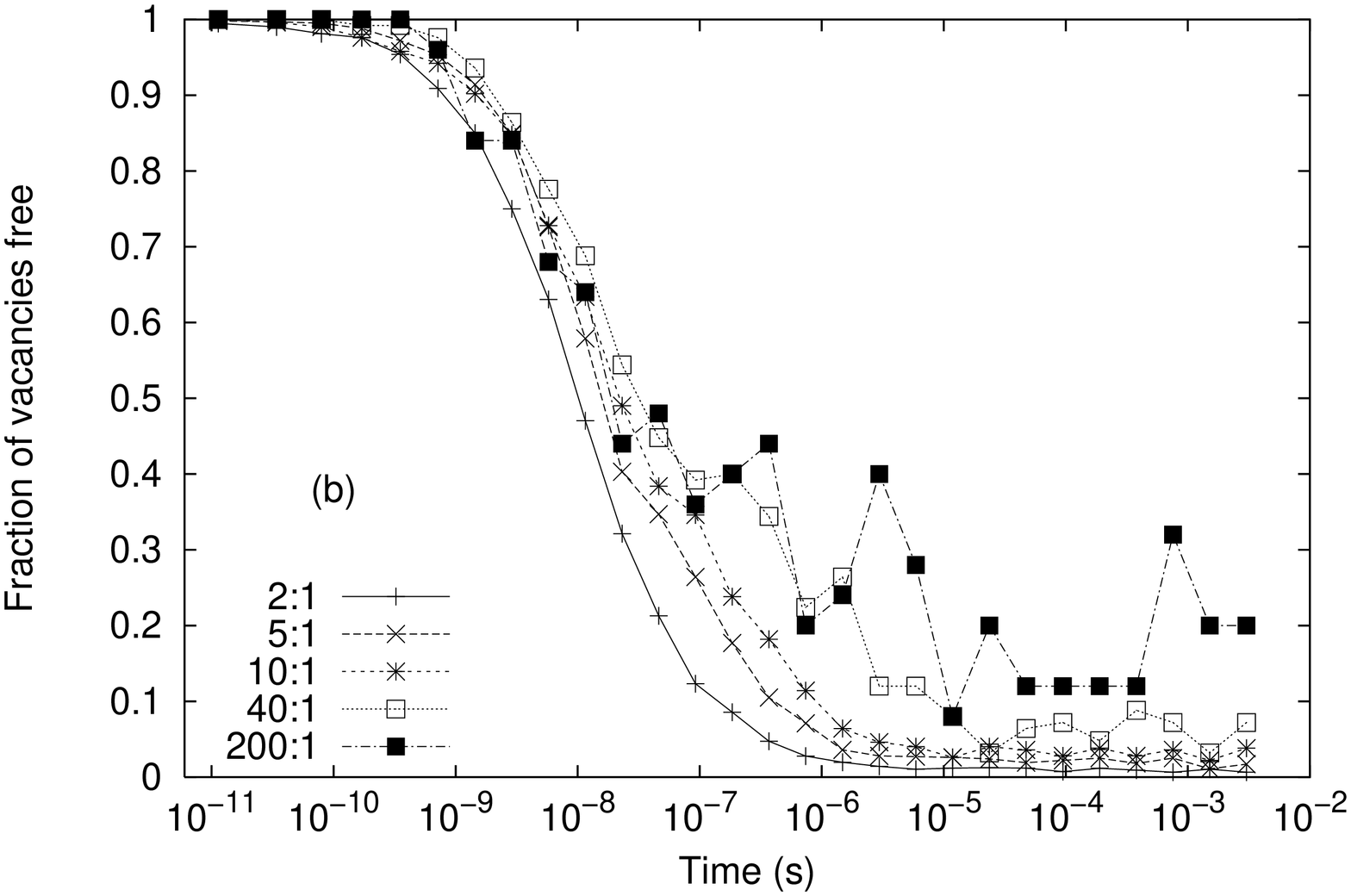}}}
\caption{\label{fig_free_ratio}
Fraction of arsenic (a) and vacancies (b) free over time for various arsenic:vacancy ratios,
with arsenic concentration $10^{19}\ \text{cm}^{-3}$ and 18 shell interactions at 900 K.}
\end{figure}
Figure \ref{fig_free_ratio} supports this trend.  With fewer vacancies in the system,
a greater fraction of both arsenic and vacancies will be free over time.
\begin{figure}\scalebox{0.5}[0.5]{\rotatebox{270}{
\includegraphics{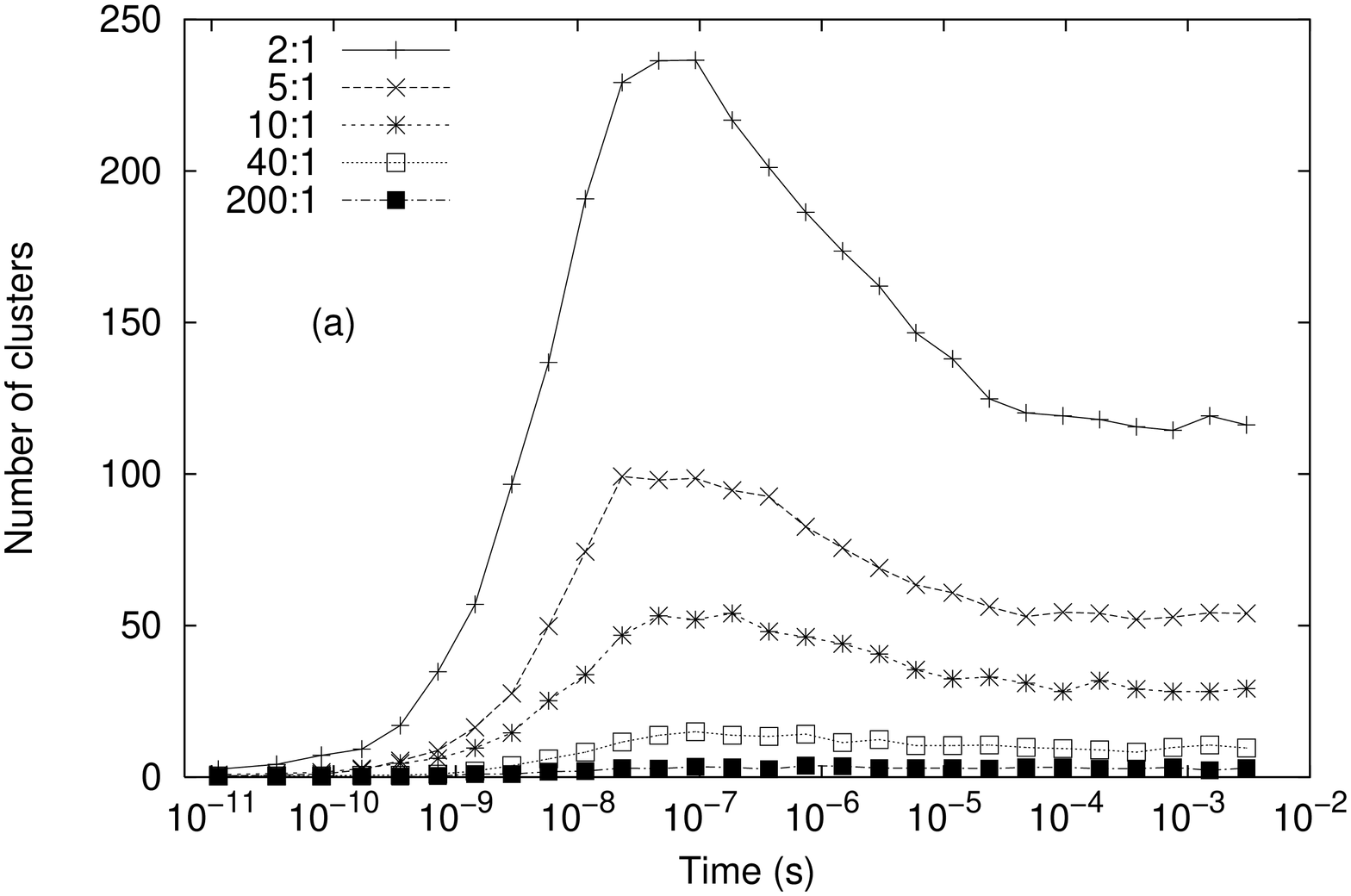}\includegraphics{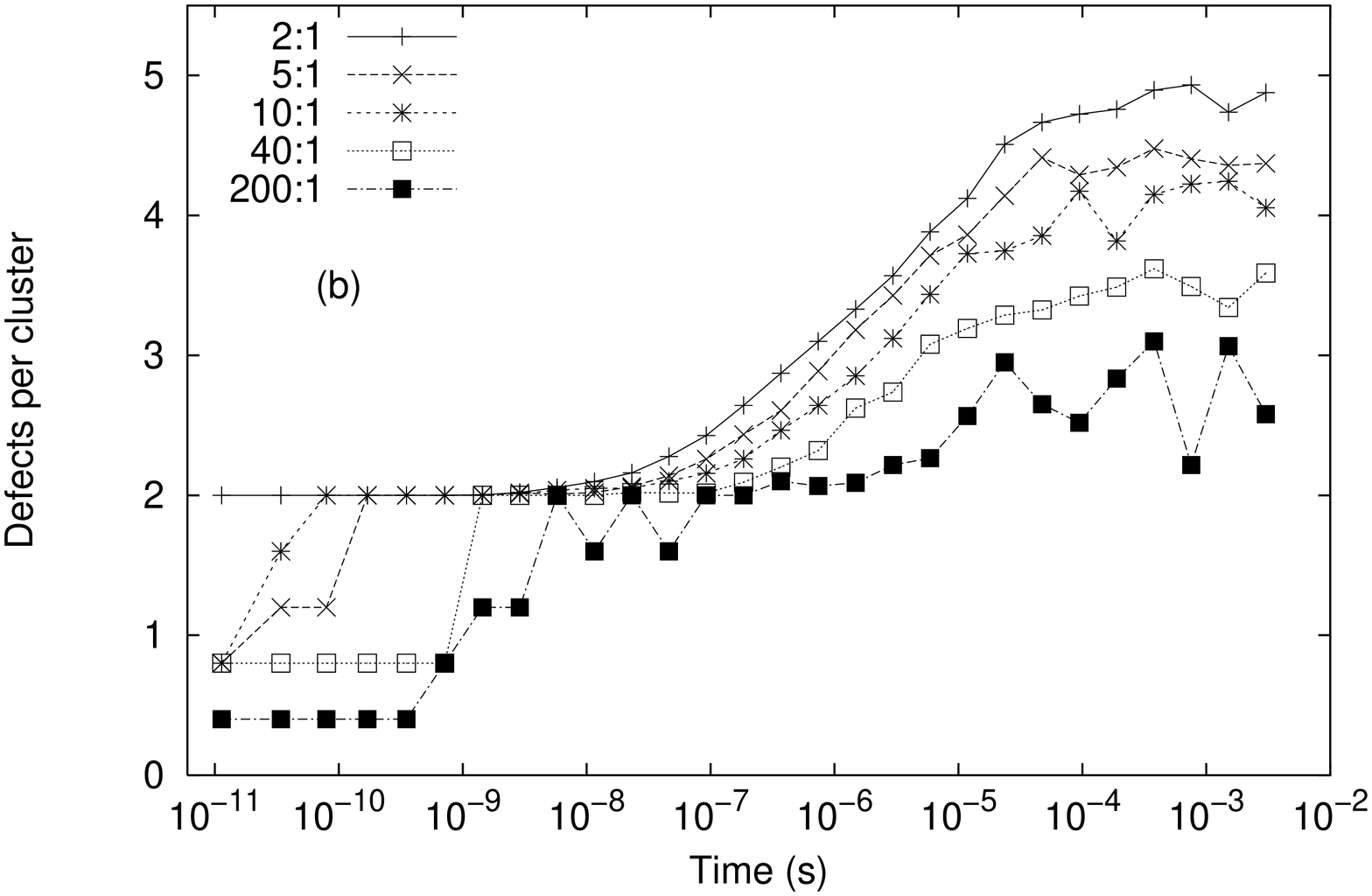}}}
\caption{\label{fig_ndefs_ratio}
Number of clusters (a) and defects per cluster (b) over time for various arsenic:vacancy 
ratios,
with arsenic concentration $10^{19}\ \text{cm}^{-3}$ and 18 shell interactions at 900 K.}
\end{figure}
\begin{figure}\scalebox{0.5}[0.5]{\rotatebox{270}{
\includegraphics{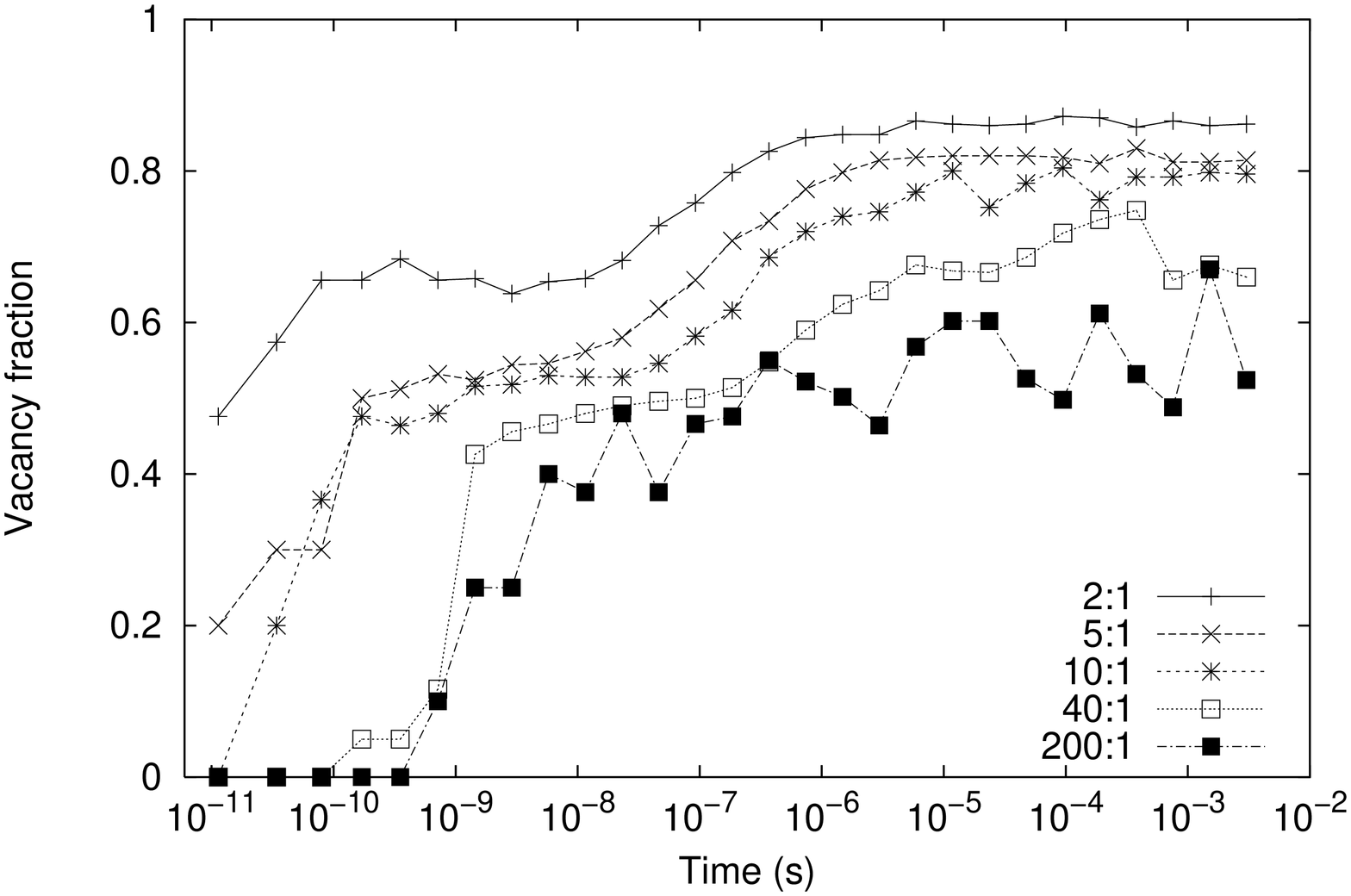}}}
\caption{\label{fig_fractV_ratio}
Vacancy fraction of defects in clusters over time for various arsenic:vacancy ratios,
with arsenic concentration $10^{19}\ \text{cm}^{-3}$ and 18 shell interactions at 900 K.}
\end{figure}
In Figure \ref{fig_ndefs_ratio} we see that a greater number of vacancies produces both a 
higher number of clusters and larger clusters, and Figure \ref{fig_fractV_ratio} shows that 
the vacancy fraction of defects in clusters will increase with vacancy concentration.  
Thus, having more vacancies in the system leads to more clusters, larger clusters, and 
more vacancies in the clusters.  This is most likely due to the strong vacancy-vacancy 
interaction at short range.

\section{Conclusion}
We have presented results of kinetic lattice Monte Carlo simulations using long range
interactions, to the 18th nearest lattice neighbor, and large numbers of defects, 
including 1000 arsenic atoms and a high vacancy concentration, which would be found 
shortly after implantation of dopant atoms into silicon.
We demonstrated that the trend at higher temperatures is toward fewer clusters 
with more defects.   At all investigated temperatures, the
vacancy fraction of defects in clusters increases over time.   At low temperatures, this
implies the dissociation of some arsenic-vacancy pairs (AsV), after which the newly free 
vacancy attaches to a nearby AsV.  At higher temperatures, mobile AsV attract free 
vacancies.  In both cases, a diffusion model based only on arsenic-vacancy pairs may be 
inadequate because of time dependent clustering effects.
\par
We also demonstrated that considering only short interaction ranges, less than 18 shells, 
produces mostly vacancy clusters, due to the strong short range attraction between 
vacancies.  This behavior manifests itself in decreasing diffusivities over time, 
exclusion of arsenic from clusters as AsV dissociate, decreasing number of clusters while 
cluster size grows, and vacancy fraction of defects in clusters approaching 1 over time.
Such behavior implies that little or no arsenic diffusion is occurring and is in 
contradiction with experimental evidence.  Therefore, long range interactions seem crucial 
for realistic simulations of defect migration in silicon.
\par
We showed that arsenic diffusivity increases for arsenic concentrations up to 
$10^{19}\ \text{cm}^{-3}$
and then decreases as arsenic concentration increases.  This behavior is due to the 
formation of a greater number of clusters at higher arsenic concentration; the size of the 
clusters does not increase substantially.  The vacancy fraction, however, changes 
significantly, transitioning from vacancy dominated clusters to arsenic dominated clusters 
at high arsenic concentrations, keeping the arsenic:vacancy ratio constant.
\par
Lastly we studied diffusion and clustering behavior as a function of the number of 
vacancies in the system.  We concluded that higher concentrations of vacancies lead to 
larger clusters,
more clusters, and more vacancies per cluster.  While a large vacancy concentration is
expected immediately after ion implantation, this may not be realistic at later
times.  More work is needed to incorporate interstitial defects, especially
silicon self interstitials so that Frenkel pair recombination can change the vacancy
concentration over time.

\begin{acknowledgments}  
We thank Keith M.\ Beardmore for contributing the VASP interaction 
potentials, Lin-Wang Wang and Andrew Canning of the Lawrence Berkeley National Laboratory 
for the PEtot code and advice on the calculations, and Mike Masquelier and Wolfgang Windl 
for useful discussions in the early stages of the work.  Initial part of the work was
supported by the Computational Nanoscience Group of Motorola, Inc.
\end{acknowledgments}



\end{document}